\documentclass[twocolumn,secnumarabic,amssymb, amsmath, nobibnotes, aps, prc]{revtex4-2}

\usepackage{color}
\usepackage{graphicx}
\usepackage{dcolumn}
\usepackage{bm}
\usepackage{hyperref}
\usepackage{mathrsfs}
\usepackage{multirow}
\usepackage{rotating}
\usepackage{url}
\usepackage{ulem}
\usepackage{enumitem}
\usepackage{diagbox}
\usepackage{threeparttable}
\usepackage{booktabs}
\usepackage[thicklines]{cancel}
\usepackage{epstopdf}
\usepackage{comment}
\usepackage{soul}

\begin{document}

	\title{Charmed hypernuclei within density-dependent relativistic mean-field theory}

	\author{Wei Yang}
	\affiliation
	{MOE Frontiers Science Center for Rare Isotopes, Lanzhou University, Lanzhou 730000, China}
	\affiliation
	{School of Nuclear Science and Technology, Lanzhou University, Lanzhou 730000, China}
	
	\author{Shi Yuan Ding}
	\affiliation
	{MOE Frontiers Science Center for Rare Isotopes, Lanzhou University, Lanzhou 730000, China}
	\affiliation
	{School of Nuclear Science and Technology, Lanzhou University, Lanzhou 730000, China}
	
	\author{Bao Yuan Sun\footnote{
			Corresponding author (Email: sunby@lzu.edu.cn)}}
	\affiliation
	{MOE Frontiers Science Center for Rare Isotopes, Lanzhou University, Lanzhou 730000, China}
	\affiliation
	{School of Nuclear Science and Technology, Lanzhou University, Lanzhou 730000, China}

\begin{abstract}
The charmed $ \Lambda_{c}^{+} $ hypernuclei are investigated within the framework of the density-dependent relativistic mean-field (DDRMF) theory. Starting from the empirical hyperon potential in symmetric nuclear matter, obtained through microscopic first-principle calculations, two sets of $\Lambda_c N$ effective interactions were derived by fitting the potentials with minimal uncertainty (Fermi momentum $k_{F,n} = 1.05~\rm{fm}^{-1}$) and near saturation density ($k_{F,n} = 1.35~\rm{fm}^{-1}$). These DDRMF models were then used to explore the $\Lambda_{c} N$ effective interaction uncertainties on the description of hypernuclear bulk and single-particle properties. A systematic investigation was conducted on the existence of bound $\Lambda_{c}^{+}$ hypernuclei. The dominant factors affecting the existence and stability of hypernuclei were analyzed from the perspective of the $\Lambda_{c}^{+}$ potential. It is found that the hyperon potential is not only influenced by the Coulomb repulsion, but by an extra contribution from the rearrangement terms due to the density dependence of the meson-baryon coupling strengths. Therefore, the rearrangement term significantly impacts the stability description for light hypernuclei, while for heavier hypernuclei, the contribution from Coulomb repulsion becomes increasingly significant and eventually dominant. The discussion then delves into the bulk and single-particle properties of charmed hypernuclei using these models. It is found that even when different models yield similar hyperon potentials for nuclear matter, different treatments of nuclear medium effects could lead to disparities in the theoretical description of hypernuclear structures. This study indicates that constraints on the $ \Lambda_{c} N $ interaction at finite densities are crucial for the study of $ \Lambda_{c}^{+} $ hypernuclear structures.
\end{abstract}
\maketitle

\section{Introduction}
Since the discovery of hyperons in the early 1950s, particles containing strange quarks have attracted significant attention from both experimental and theoretical physicists \cite{Danysz1953.44.350}. Hyperons, which carry new degrees of freedom beyond nucleons, are free from the nucleon's Pauli exclusion principle in nuclei, and then, similar to impurities, hyperons can deep into the nucleus and combine with nucleons to form hypernuclear systems. These properties make hyperons sensitive probes for studying nuclear structure and specific nuclear features. Research on hyperons within nuclei helps us understand the baryon-baryon interactions in nuclear matter and their impact on nuclear properties \cite{Yamamoto-Prog.Theo.Phys.Suppl117(1994)361, B.F.Gibson-PR257(1995)349}. This knowledge is also crucial for understanding the matter in neutron stars, where hyperons may appear \cite{Prakash1997Phys.Rep280.1, Lattimer2004Science304.536, Tolos2020PPNP112.103770}. So far, a wealth of experimental data on hypernuclei with strangeness $ S=-1 $ ($ \Lambda $ and $ \Sigma $ hypernuclei) has been detected, as well as a few hypernuclei with strangeness $ S=-2 $ ($ \Lambda\Lambda $ and $ \Xi $ hypernuclei) \cite{Aoki1993PTP89.493, Takahashi2001PRL87.212502, Yamaguchi2001PTP105.627, Hashimoto2006PPNP57.564, Nakazawa2010NPA835.207, Feliciello2015Rep.Prog.Phys78.096301, Gal2016Rev.Mod.Phys88.035004}. Based on this experimental information, extensive theoretical research has been conducted on the structure of hypernuclei and neutron star matter containing hyperons \cite{SCHAFFNER1994Ann.Phys235.35, Cugnon2000PRC62.064308, H.Shen2002NPA707.469, SunTT2016PRC94.064319, Qian.Zhuang-PRC106(2022)054311, ShiYuan.Ding-CPC47(2023)124103}. In addition to the above-mentioned hyperons containing strange quarks, theory has also predicted a particle containing charmed quark, whose composition is very similar to that of the $ \Lambda $ hyperon. It can be viewed as the strange quark in the $ \Lambda $ hyperon replaced by a charmed quark, and was experimentally evidented in the early 1970s \cite{Cazzoli.E.G-PRL34(1975)1125, Knapp.B-PRL37(1976)882}. Due to their similar composition, charmed particles may exhibit similar behaviors to $ \Lambda $ hyperons, such as moving deep into the nucleus to form charmed hypernuclei, providing us with another unique perspective for studying nuclear structure. Research on the charmed hypernuclei may be a limited source of information to help us understand the charmed baryon-nucleon interactions within the SU(4) symmetry framework, and the extraction of these interactions also helps us to test theoretical model predictions such as dynamical chiral symmetry breaking (DCSB) \cite{A.Hosaka-PPNP96(2017)88, Krein2019APIConProce2130.020022}.

Experimental detection provides a direct and effective approach to study hypernuclear structure. In the past, it was suggested that charmed hypernuclei could be produced through the charm exchange reaction, namely the ($ D $, $ \pi $) reaction \cite{T.Bressani-Il.Nuovo.Cimento.A102(1989)597, Bunyatov.S.A-Il.Nuovo.Cimento.A104(1991)1361}. Due to the extremely short lifetimes and high momenta of $ D $ mesons, they are difficult to be captured by nucleons in these reactions, which poses significant challenges for the formation and study of charmed hypernuclei. Consequently, over the past few decades, only a few possible signals of charmed hypernuclei have been detected at Dubna \cite{Batusov.Yu.A-JETPLett33(1981)52, LYUKOV.VV-Il.Nuovo.Cimento.A102(1989)583}. In recent years, another effective method for producing charmed hypernuclei, namely the antiproton-nucleus collisions method, has been proposed. This method does not require the production of additional $ D $ mesons, thereby increasing the possibility of forming charmed hypernuclei \cite{U.Wiedner-PPNP66(2011)477}. The $ \bar{\mathrm{P}} $ANDA experiment at FAIR (GSI) is expected to study the production of charmed hypernuclei using the new method \cite{U.Wiedner-PPNP66(2011)477}, and the experimental feasibility was also theoretically analyzed\cite{R.Shyam-PLB770(2017)236}. Furthermore, the J-PARC facility, which can provide high-intensity and high-momentum proton beams, also offers an ideal experimental platform for the production of charmed hypernuclei \cite{Riedl.J-EPJC52(2007)987, H.Fujioka-arXiv1706.07916(2017)}. With the gradual construction and upgrading of facilities for radioactive ion beams, accurate information about the charmed hypernuclei is expected to be obtained in the future, providing strong support for the study of hypernuclear structures.

Another effective approach to studying hypernuclear structures is through theoretical analysis, and extensive research has been conducted based on various theoretical models. Since the $ \Lambda_{c}^{+} $ particle carries the same positive charge as the proton, its Coulomb interaction with protons in the nucleus affects the stability of $ \Lambda_{c}^{+} $ hypernuclei, making it challenging to form a hypernucleus by binding with nucleons. Consequently, after the experimental observation of charmed particles, theoretical physicists have primarily focused on discussing the existence and stability of $ \Lambda_{c}^{+} $ hypernuclei \cite{Cazzoli.E.G-PRL34(1975)1125}. In 1977, the possible existence of $ \Lambda_{c}^{+} $ hypernuclei was first explored theoretically based on SU(4) symmetry \cite{Dover.C.B-PRL39(1977)1506}. Subsequently, many theoretical models have been extended to include the $ \Lambda_{c}^{+} $ degree of freedom, systematically investigating $ \Lambda_{c}^{+} $ hypernuclei from light to heavy. For lighter hypernuclei, systematic studies have been conducted using few-body methods based on cluster models and Faddeev equations \cite{Bhamathi.G-PRC24(1981)1816, Bando.H-PLB109(1982)164, Gibson.B.F-PRC27(1983)2085, Garcilazo.H-PRC92(2015)024006}. Recently, in-depth discussions on the existence of the $^3_{\Lambda_{c}}$H hypernucleus have been conducted based on the quark-delocalization color-screening (QDCSM) model \cite{WuSiyu-PRC109(2024)014001}. Density functional theory (DFT) is undoubtedly an ideal approach for studying medium and heavy hypernuclei, as it can describe the single-particle and collective properties of finite nuclei in the almost entire nuclear chart. In previous work, several density functional theory approaches have been extended to research the structure of $ \Lambda_{c}^{+} $ hypernuclei, such as the quark meson coupling (QMC) model \cite{Tsushima.K-PRC67(2003)015211, R.Shyam-PLB770(2017)236}, the quark mean-field (QMF) model \cite{Wu.Linzhuo-PRC101(2020)024303}, the Skyrme-Hartree-Fock (SHF) approach \cite{Guven-PRC104(2021)064306, Liu.Yi.Xiu-PRC108(2023)064312}, and the relativistic mean-field (RMF) approach \cite{Tan.Yu.Hong-PRC70(2004)054306, Yu.Hong.Tan-Europhys.Lett67(2004)355}. In addition to DFT, the perturbation many-body approach based on the nuclear matter G-matrix has also seen further development recently, achieving a unified and self-consistent description of $ \Lambda_{c}^{+} $ hypernuclear structures from light to heavy \cite{Vidana.I-PRC99(2019)045208, Haidenbauer.J-EPJA56(2020)195}. Based on these theoretical models, detailed analyses have been carried out on the existence and stability of charmed hypernuclei, and the impurity effects induced by the introduction of $ \Lambda_{c}^{+} $ particles have been further investigated. These theoretical analyses provide valuable guidance for experiments on the production and detection of charmed hypernuclei.

Despite extensive theoretical analyses of the $ \Lambda_{c}^{+} $ hypernuclear structure, the lack of experimental information has made it difficult to construct the $ \Lambda_{c} N $ interaction from a unified starting point, which has also led to additional uncertainties in the results of various theoretical models for the $ \Lambda_{c}^{+} $ hypernuclear structure. In early works, pivotal information on the $ \Lambda_{c} N $ interaction, such as low-energy scattering parameters and hyperon potentials, was directly obtained based on SU(4) symmetry \cite{Dover.C.B-PRL39(1977)1506, Bhamathi.G-PRC24(1981)1816, N.I.Starkov-NPA450(1986)507}. Some studies have attempted to construct the $ \Lambda_{c} N $ interaction by scaling the $ \Lambda_{c}^{+} $ hyperon potential to that of the $ \Lambda $ hyperon or using other specific values, but the significant differences in the constructed interactions have also introduced large uncertainties in the theoretical analysis of hypernuclear structure \cite{Yu.Hong.Tan-Europhys.Lett67(2004)355, Liu.Yi.Xiu-PRC108(2023)064312}. Recently, lattice QCD simulations have explored the $ \Lambda_{c} N $ interaction under different unphysical pion masses. Subsequent work used chiral effective field theory to extrapolate the simulation results, obtaining the effective $ \Lambda_{c} N $ interaction at a pion mass of $ m_{\pi} = 138 $ MeV \cite{Haidenbauer.J-EPJA54(2018)199}. Building upon this, the extrapolated results including the $ \Sigma_{c} N $ channel have been studied, and it was found that the $ \Sigma_{c} N $ coupling has little impact on the $ \Lambda_{c} N $ interaction at low energies \cite{Haidenbauer.J-EPJA56(2020)195}. Additionally, the properties of $ \Lambda_{c}^{+} $ hypernuclear bound states and resonance states have also been explored by considering different symmetries in the effective Lagrangian \cite{Liu.Yan.Rui-PRD85(2012)014015, Maeda.Saori-PRC98(2018)035203}. Based on the derived two-body $ \Lambda_{c} N $ interaction, various theoretical models have been employed to obtain information on $ \Lambda_{c}^{+} $ hyperon in nuclear matter. For example, analyses based on SU(4) symmetry indicate that the $ \Lambda_{c}^{+} $ hyperon potential in nuclear matter ranges from -20 MeV to -28 MeV \cite{R.Gatto-Il.Nuovo.Cimento.A46(1978)313, Bhamathi.G-PRC24(1981)1816, Bando.H-Prog.Theor.Phys.Suppl81(1985)197, N.I.Starkov-NPA450(1986)507}. Recent lattice QCD simulations and their extrapolated results suggest a hyperon potential depth of less than 20 MeV \cite{T.Miyamoto-NPA971(2018)113, Yasui.Shigehiro-PRC100(2019)065201, Haidenbauer.J-EPJA56(2020)195}, consistent with results obtained using parity-projected QCD sum rules, which propose an attractive potential of $ U_{\Lambda_{c}} \sim -20 $ MeV at normal nuclear density \cite{Ohtani2017PRC96.055208}. The $ \Lambda_{c} N $ interaction is a key aspect of the theoretical analysis of hypernuclear structure, significantly affecting the description of hypernuclear properties. Therefore, selecting a reasonable $ \Lambda_{c} N $ interaction is crucial to ensure more reliable results.

In addition to the $ \Lambda_{c} N $ interaction, a reliable description of hypernuclear structure also relies on the adopted many-body methods. As a significant branch of density functional theory, RMF theory maintains theoretical superiority, which is capable of describing not only infinite nuclear matter but the single-particle and collective properties of finite nuclei in the almost entire nuclear chart, and has been extended to the study of strangeness degrees of freedom \cite{Brockmann-PLB69(1977)167, Bouyssy1981PLB99.305, Glendenning1991PRL67.2414, Jennings.B.K-PRC49(1994)2472, Sugahara.Yuichi-Prog.Theo.Phys.92(1994)803, Vretenar1998PRC57.R1060, Hu2014PRC90.014309, Wu2017PRC95.034309, Liu2018PRC98.024316, ShiYuan.Ding-CPC47(2023)124103}. Considering the existence and stability issues of the $ \Lambda_{c}^+ $ hypernuclei in the large mass range we are currently concerned with, the RMF theory is undoubtedly the ideal research method. In fact, RMF theory based on nonlinear coupling has already been extended to study the structure of $ \Lambda_{c}^{+} $ hypernuclei. However, due to the lack of experimental information, the construction of the $ \Lambda_{c} N $ interaction lacks a reasonable basis, resulting in significant uncertainties in the results \cite{Tan.Yu.Hong-PRC70(2004)054306, Yu.Hong.Tan-Europhys.Lett67(2004)355}. Therefore, it is necessary to start from a more reasonable $ \Lambda_{c} N $ interaction and combine it with RMF theory to achieve a reliable description of hypernuclear structure. Additionally, considering that the hyperon is located within the nucleus, the $ \Lambda_{c} N $ interaction will be significantly affected by medium effects, and different treatments for medium effects need to be carefully discussed regarding their impact on the bulk and single-particle properties of $ \Lambda_{c}^{+} $ hypernuclei. Microscopic calculations based on the Dirac-Brueckner-Hartree-Fock (DBHF) theory indicate that nuclear medium effects have a significant impact on the description of nuclear structure \cite{Brockmann.R-PRL68(1992)3408}. By treating the meson-nucleon coupling strengths as functions of the baryon density, nuclear medium effects can be effectively considered. The density-dependent relativistic mean-field (DDRMF) and density-dependent relativistic Hartree-Fock (DDRHF) theories, developed based on this idea, achieve a self-consistent and unified description of different nuclear medium densities by introducing density-dependent meson-nucleon coupling strengths \cite{Brockmann.R-PRL68(1992)3408, Wen.Hui.Long-PLB640(2006)150}. Numerous related studies have been conducted in dense matter and finite nuclei based on these methods. For instance, they have explored nuclear symmetry energy \cite{Ring.P-PhysRevC.77(2008)034302, Sun2008PRC78.065805, Roca.Maza-PRC84(2011)054309, Long.W.H-PRC85(2012)025806, Zhao2015JPG42.095101, Liu.Z.W-PRC.97(2018)025801}, nucleon effective mass \cite{Wen.Hui.Long-PLB640(2006)150, Li2016PRC93.015803}, liquid-gas phase transition \cite{G.H.Zhang-PLB.720(2013)148-1152, Yang.S-PRC.100(2019)054314, Yang.S-PRC.103(2021)014304}, equation of state (EOS) of dense matter \cite{Shen.G-PhysRevC.82(2010)015806, Long.W.H-PRC85(2012)025806, Fortin2020PRD101.034017}, neutron star \cite{Hofmann-PhysRevC.64(2001)025804, Avancini.S.S-PhysRevC.75(2007)055805, WangSha-PhysRevC.90(2014)055801, Liu.Z.W-PRC.97(2018)025801}, shell evolution \cite{Long2008Euro.Lett82.12001, Wang2013PRC87.047301, Li2016PLB753.97}, neutron skin effects \cite{Avancini.S.S-PhysRevC.76(2007)064318, Avancini.S.S-PhysRevC.75(2007)055805} and novel feature in exotic nuclei \cite{Hofmann-PhysRevC.64(2001)034314, D.Vretenar-PhysRep.409(2005)101, A.V.Afanasjev-PLB.726(2013)680-684}. Additionally, density-dependent couplings fundamentally alter the balance between attraction and repulsion in nuclear forces, thereby affecting the description of finite nuclear structure and nuclear matter properties. For example, a new type of density-dependent effective interaction DD-LZ1 was proposed by adopting a unique density-dependent form, which solves the common problem of $ Z = 58 $, $ 92 $ pseudo-shell closures within the RMF framework, and demonstrates great advantages in describing the crust of neutron stars and the maximum mass of hyperonic neutron stars \cite{Yang2022PRD105.063023, Xia.Cheng.Jun-PRC105(2022)045803}. As a further extension, the DDRMF/DDRHF theories have been applied to the study of single-$\Lambda$ hypernuclear structures, with a particular focus on the impact of medium effects on hyperon spin-orbit splitting \cite{Qian.Zhuang-PRC106(2022)054311, ShiYuan.Ding-CPC47(2023)124103}. Therefore, it is worthwhile to further explore the effects of different treatments of effective nuclear forces in the medium on the description of $ \Lambda_{c}^{+} $ hypernuclear structures.

In this work, the stability and properties of $ \Lambda_c^+ $ hypernuclei are investigated using the DDRMF theory. The $ \Lambda_c N $ interaction at finite densities is determined by fitting the $ \Lambda_c^+ $ potential constraints in symmetric nuclear matter as provided by Ref. \cite{Haidenbauer.J-EPJA56(2020)195}. According to the behaviors of the constraint potential, we select the potentials at Fermi momentum $ k_{F,n} = 1.05~\rm{fm}^{-1} $ and $ k_{F,n} = 1.35~\rm{fm}^{-1} $ as the fitting targets. These two Fermi momentum values are used to study the influence of different $ \Lambda_cN $ and $ NN $ interactions on the properties of hypernuclei. Additionally, the impact of density-dependent coupling strengths on the description of hypernuclei is studied. In hypernuclear research, impurity effects due to the introduction of $ \Lambda_{c}^+ $ hyperon are also studied. This paper is organized as follows. In Sec. \ref{Theoretical Framework}, we will introduce the framework of the DDRMF model for the charmed hypernuclei. In Sec. \ref{Results and Discussion}, the determination of $ \Lambda_c N $ interaction and calculation results of the charmed hypernuclei will be discussed. Finally, a summary is provided in Sec. \ref{Summary and Outlook}.

\section{Theoretical Framework}\label{Theoretical Framework}

In this section, we briefly introduce the general formalism of the DDRMF theory incorporating the $ \Lambda_{c}^+ $ hyperon degrees of freedom, which starts from the following Lagrangian density
\begin{align}\label{eq:Lagrangian}
	\mathscr{L} = \mathscr{L}_{B}+ \mathscr{L}_{\varphi} + \mathscr{L}_{I}.
\end{align}
The terms of the free fields are given by
\begin{align}\label{eq:LagrangianNLRMF}
	\mathscr{L}_{B}=&\sum_B\bar{\psi}_B\left(i\gamma^\mu\partial_\mu-M_B\right)\psi_B,\\
	\mathscr{L}_{\varphi}=&+\frac{1}{2}\partial^\mu\sigma\partial_\mu\sigma-\frac{1}{2}m_\sigma^2\sigma^2\nonumber\\
	&-\frac{1}{4}\Omega^{\mu\nu}\Omega_{\mu\nu}+\frac{1}{2}m_\omega^2\omega^\mu\omega_\mu\nonumber\\
	&-\frac{1}{4}\vec{R}^{\mu\nu}\cdot\vec{R}_{\mu\nu}+\frac{1}{2}m_\rho^2\vec{\rho}^{\mu}\cdot\vec{\rho}_\mu\nonumber\\
	&+\frac{1}{2}\partial^\mu\vec{\delta}\partial_\mu\vec{\delta}-\frac{1}{2}m_\delta^2\vec{\delta}^2\nonumber\\
	&-\frac{1}{4}F^{\mu\nu}F_{\mu\nu},
\end{align}
where the index $ B $ (and later $ B' $) represents either a nucleon $ N $ or the charmed hyperon $ \Lambda_{c}^{+} $, and $ \Sigma_{B} $ represents the sum over nucleons ($ n, p $) and the charmed hyperon ($ \Lambda_{c}^{+} $). The masses of baryons and mesons are given by $ M_{B} $ and $ m_{\phi} $ ($ \phi = \sigma, \omega^{\mu}, \vec{\rho}^{\mu}, \vec{\delta} $), respectively. Additionally, $ \Omega^{\mu\nu}, \vec{R}^{\mu\nu} $, and $ F^{\mu\nu} $ are the field tensors of the vector mesons $ \omega^{\mu} $, $ \vec{\rho}^{\mu} $ and the photon $ A^{\mu} $, respectively. The interaction between baryons and mesons (photon) can be expressed as $ \mathscr{L}_{I} $,
\begin{align}\label{eq:Lagrangian-interaction}
	\mathscr{L}_{I} &= \sum_{B} \bar{\psi}_{B}\left(-g_{\sigma B}\sigma - g_{\omega B}\gamma^\mu\omega_{\mu}-eQ_{B}\gamma^{\mu} A_{\mu} \right)\psi_{B}\nonumber\\
	&+ \sum_{N} \bar{\psi}_{N}\left(-g_{\delta N}\vec{\tau}_{N}\cdot\vec{\delta}-g_{\rho N}\gamma^{\mu} \vec{\tau}_{N}\cdot\vec{\rho}_{\mu}\right)\psi_{N}.
\end{align}
Here $ g_{\phi B}~(g_{\phi N}) $ represents the coupling strengths for various meson-baryon channels, and $ \vec{\tau}_{N} $ is the isospin operator with the third component $ \tau_{3,N} = 1 $ for the neutron, $ \tau_{3,N} = -1 $ for the proton. To maintain the simplicity of the theoretical framework, we define the symbol $ Q_{B} $, for nucleons, $ Q_{N} = \frac{1-\tau_{3,N}}{2} $, and for $ \Lambda_{c}^+ $ hyperon, $ Q_{\Lambda_{c}} = 1 $. Within the framework of the RMF theory, an additional coupling term between hyperons and the $ \omega $-tensor is often considered to ensure that the spin-orbit splitting of hyperons such as the $ \Lambda $ aligns with experimental results \cite{Cohen.Joseph-PRC44(1991)1181, Sugahara.Yuichi-Prog.Theo.Phys.92(1994)803, Jennings.B.K-PRC49(1994)2472, Sun.T.T-PRC96(2017)044312}. In fact, due to its larger mass, the spin-orbit splitting for the $ \Lambda_{c}^{+} $ is significantly reduced compared to the $ \Lambda $, making this effect negligible for our results \cite{Liu.Yi.Xiu-PRC108(2023)064312, Wu.Linzhuo-PRC101(2020)024303, Guven-PRC104(2021)064306}. Therefore, we have disregarded this effect in our work.

In the density-dependent relativistic mean-field theory, the meson-baryon (nucleon) coupling strength is treated as a function of the baryon density $ \rho_{b} $. This approach phenomenologically incorporates the nuclear in-medium effects. Specifically, for the isoscalar mesons ($ \sigma $ and $ \omega^{\mu} $), their coupling strengths with baryons can generally be expressed as
\begin{align}\label{eq:coupling_constants1}
	g_{\phi B}\left(\rho_{b}\right)&=g_{\phi B}(0) a_{\phi B}\frac{1+b_{\phi B}(\xi+d_{\phi B})^2}{1+c_{\phi B}(\xi+e_{\phi B})^2},
\end{align}
where $ \xi = \rho_{b}/\rho_{0} $ with $ \rho_{0} $ being the saturation density of the nuclear matter, and $ g_{\phi B}(0) $ are the coupling strengths at $ \rho_{b} = 0 $. It is worth noting that, aside from the effective interaction DD-ME$ \delta $ \cite{Roca.Maza-PRC84(2011)054309}, the other Lagrangians used in this work have $ e_{\phi B} $ equal to $ d_{\phi B} $ in Eq. \eqref{eq:coupling_constants1}. Furthermore, for the isovector mesons ($ \vec{\delta} $ and $ \vec{\rho}^{\mu} $), the effective interaction DD-ME$ \delta $ uses a form consistent with that of the isoscalar mesons, whereas the other effective interactions follow an exponential decay form, which can be expressed as
\begin{align}
	g_{\phi B}\left(\rho_{b}\right)&=g_{\phi B}(0) e^{-a_{\phi B} \xi}.
\end{align}

Based on the Lagrangian density $ \mathscr{L} $ of Eq. \eqref{eq:Lagrangian}, the effective Hamiltonian operator for the $ \Lambda_{c}^{+} $ hypernucleus can be derived through the general Legendre transformation. It can be written as follows
\begin{align}\label{eq:Hamiltonian}
	\hat{H}&=\int dx \sum_{B} \bar{\psi}_{B}(x)(-i\boldsymbol{\gamma }\cdot \boldsymbol{\nabla}+M_{B })\psi_{B}(x)\nonumber\\
	&+\frac{1}{2} \int dx \sum_{BB'} \sum_{\varphi} \left[\bar{\psi}_{B} \mathscr{G}_{\varphi B} \psi_{B}\right]_{x} D_{\varphi}(x,x') \left[\bar{\psi}_{B'} \mathscr{G}_{\varphi B'} \psi_{B'}\right]_{x'},
\end{align}
where $ x $ is the four-vector $ (t, \bm{x}) $ and $ \varphi = \sigma, \omega^{\mu}, \vec{\rho}^{\mu}, \vec{\delta}, A^{\mu} $. The interaction vertices for various mesons (photon)-baryon coupling channels are denoted as $ \mathscr{G}_{\varphi B}(x) $, and $ D_{\varphi} $ are defined as the meson (photon) propagators \cite{Qian.Zhuang-PRC106(2022)054311}. Considering the simplicity of the theoretical framework, only the form for the $ \Lambda_{c}^{+} $ will be provided, with the specifics of the nucleon part detailed in Refs. \cite{Qian.Zhuang-PRC106(2022)054311, ShiYuan.Ding-CPC47(2023)124103}. Since the $ \Lambda_{c}^{+} $ is a charged particle with zero isospin, it only participates in interactions with isoscalar mesons ($ \sigma $ and $ \omega^{\mu} $) and photon. Consequently, the interaction vertices of the $ \Lambda_{c}^{+} $ hyperon with various mesons (photon) are given by
\begin{subequations}\label{eq:vertice for sigmaomega}
	\begin{align}
		\mathscr{G}_{\sigma \Lambda_{c}}(x) = &+g_{\sigma \Lambda_{c}}(x),\\
		\mathscr{G}_{\omega \Lambda_{c}}^\mu(x) = &+g_{\omega \Lambda_{c}}(x)\gamma^{\mu},\\
		\mathscr{G}_{A \Lambda_c}^\mu(x) = &+eQ_{\Lambda_{c}}\gamma^{\mu}.
	\end{align}
\end{subequations}

Under the no-sea approximation, the hyperon field operator $ \psi_{\Lambda_{c}} $ can be expanded using the positive energy solutions
\begin{align}
	\psi_{\Lambda_{c}}(x)
	&=\sum_if_i(\bm{x})e^{-i\epsilon_i t}c_i.\label{eq:fi}
\end{align}
Here, $ f_{i}(\bm{x}) $ denotes the Dirac spinor, and $ c_{i} $ represents the annihilation operator for state $ i $. The energy functional $ E $ of the hypernuclear system can be obtained by taking the expectation value of the Hamiltonian operator with respect to the Hartree-Fock ground state $ |\Phi_0\rangle $.

Under the spherical approximation, the Dirac spinor $ f_{i}(\bm{x}) $ of the $ \Lambda_{c}^{+} $ hyperon is expanded in the following form
\begin{align}
	f_{n\kappa m}(\bm{x}) =  \frac{1}{r} \left(\begin{array}{c}iG_{a,\Lambda_{c}}(r)\Omega_{\kappa m}(\vartheta,\varphi)\\ F_{a,\Lambda_{c}}(r)\Omega_{-\kappa m}(\vartheta,\varphi) \end{array}\right).
\end{align}
Here, the index $ a $ comprises a set of good quantum numbers $ (n\kappa) = (njl) $, with $ \Omega_{\kappa m} $ is the spherical spinor. Correspondingly, the propagator in Eq. \eqref{eq:Hamiltonian} can be expanded in terms of spherical Bessel ($ R^{LL} $) and spherical harmonic ($ Y_{LM} $) functions as
\begin{align}\label{eq:propagator}
	D_{\varphi}(\bm{x},\bm{x}^{\prime}) = \sum_{L=0}^{\infty}\sum_{M=-L}^{L}(-1)^{M}R^{\varphi}_{LL}\left( r, r^{\prime}\right) Y_{LM}\left(\bm{\Omega}\right)Y_{L-M}\left(\bm{\Omega}^{\prime}\right),
\end{align}
where $ \bm{\Omega} = (\vartheta,\varphi) $, and $ R_{LL} $ contains the modified Bessel functions $ I $ and $ K $ \cite{abramowitz1964handbook, Varshalovich_quantum_1988}.

The single-particle properties of the $ \Lambda_{c}^{+} $ hyperon can be determined by solving the Dirac equation,
\begin{align}\label{eq:Dirac}
	\varepsilon_{a,\Lambda_{c}}
	\begin{pmatrix}
		G_{a,\Lambda_{c}}(r) \\ F_{a,\Lambda_{c}}(r)
	\end{pmatrix} =&
	\begin{pmatrix}
		\Sigma_+^{\Lambda_{c}}(r) &~ \displaystyle-\frac{d}{dr}+\frac{\kappa_{a,\Lambda_{c}}}{r} \\
		\displaystyle\frac{d}{dr}+\frac{\kappa_{a,\Lambda_{c}}}{r} &~ -\left[2M_{\Lambda_{c}}-\Sigma_-^{\Lambda_{c}}(r)\right]
	\end{pmatrix}\nonumber\\
	&\begin{pmatrix}
		G_{a,\Lambda_{c}}(r) \\ F_{a,\Lambda_{c}}(r)
	\end{pmatrix}.
\end{align}
Here, the self-energies $ \Sigma_\pm^{\Lambda_{c}} = \Sigma_{0, \Lambda_{c}}\pm\Sigma_{S, \Lambda_{c}} $ of $ \Lambda_{c}^{+} $ hyperon are composed by the vector and scalar terms. The scalar self-energy $ \Sigma_{S, \Lambda_{c}} = \Sigma_{S, \Lambda_{c}}^{\sigma} $, and the time component of the vector one is given by
\begin{align}
	\Sigma_{0,\Lambda_{c}}(r) = \sum_{\varphi}\Sigma_{0,\Lambda_{c}}^{\varphi}(r)+\Sigma_{R}(r),
	\label{eq:Sig0}
\end{align}
where $ \varphi = \omega^{\mu}, A^{\mu} $ for $ \Lambda_{c}^{+} $ hyperon. Specifically, the contributions to the self-energy from isoscalar mesons ($ \sigma $ and $ \omega^{\mu} $) as well as photon can be expressed as:
\begin{subequations}
	\begin{align}
		\Sigma_{S,\Lambda_{c}}^{\sigma}(r)&=-g_{\sigma \Lambda_{c}}(r)\sum_{B}\int r^{\prime2}dr^{\prime} g_{\sigma B}(r^{\prime})\rho_{s,B}(r^{\prime})R^{\sigma}_{00}(r,r^{\prime}),\label{eq:Sigma_S,B}\\
		\Sigma_{0,\Lambda_{c}}^{\omega}(r)&=+g_{\omega \Lambda_{c}}(r)\sum_{B}\int r^{\prime2}dr^{\prime} g_{\omega B}(r^{\prime})\rho_{b,B}(r^{\prime})R^{\omega}_{00}(r,r^{\prime}),\label{eq:Sigma_0,B}\\
		\Sigma_{0,\Lambda_{c}}^{A}(r)&=+e\sum_{B}\int r^{\prime2}dr^{\prime} e\rho_{b,B}(r^{\prime})Q_{B}R_{00}^{A}(r,r^{\prime}).
	\end{align}
\end{subequations}
Here, $ \rho_{s, B} $ and $ \rho_{b, B} $ represent the scalar and baryon densities, respectively.

In the DDRMF framework, to ensure self-consistency between the energy density functional and single-particle properties, additional rearrangement terms are introduced into the baryon self-energies due to the density-dependent coupling strengths \cite{Typel.S-NPA656(1999)331}. For the $ \Lambda_{c}^{+} $ hypernucleus, the rearrangement term $ \Sigma_{R} $ includes contributions from both nucleon and hyperon. For nucleon, contributions from various coupling channels need to be considered, whereas for hyperon, the contribution arises solely from the isoscalar coupling channels. Here, for the sake of simplicity, only the contributions from the $ \Lambda_{c}^{+} $ hyperon are provided,
\begin{align}
	\Sigma_{R}^{\Lambda_{c}}(r) & = \frac{1}{g_{\sigma \Lambda_c}} \dfrac{\partial g_{\sigma \Lambda_c}}{\partial \rho_{b}}\rho_{s,\Lambda_c}\Sigma_{S,\Lambda_c}^{\sigma}(r)\notag \\ & + \frac{1}{g_{\omega \Lambda_c}} \dfrac{\partial g_{\omega \Lambda_c}}{\partial \rho_{b}}\rho_{b,\Lambda_c}\Sigma_{0,\Lambda_c}^{\omega}(r).
	\label{eq:Erea}
\end{align}

\section{Results and Discussion}\label{Results and Discussion}

In this section, we extend the DDRMF theory to include the degrees of freedom of $ \Lambda_{c}^{+} $ hyperon. By combining the results of microscopic model calculations, we construct an effective $ \Lambda_{c} N $ interaction. Based on this, we conduct a thorough discussion on the existence and stability of $ \Lambda_{c}^{+} $ hypernuclei, as well as explore the description of bound hypernuclear bulk and single-particle properties. To effectively account for the impact of nuclear medium effects on the description of hypernuclear structures, we select several sets of density-dependent Lagrangians for $ NN $ effective interactions. These include TW99 \cite{S.Typel-NPA656(1999)331}, PKDD \cite{Long.Wenhui-PRC69(204)034319}, DD-LZ1 \cite{Bin.Wei-CPC44(2020)074107}, DD-ME2 \cite{Lalazissis.G.A-PRC71(2005)024312}, DD-MEX \cite{A.Taninah-PLB800(2020)135065, Rather.Ishfaq.A-PRC103(2021)055814}, and DD-ME$ \delta $ \cite{Roca.Maza-PRC84(2011)054309}. Among these, the effective interaction DD-ME$ \delta $ includes a more comprehensive contribution of meson-nucleon coupling channels by introducing the isovector scalar $ \delta $ meson. Specifically, the Dirac equation is solved in a radial box of size $ R = 20 $ fm with a step of $ 0.1 $ fm. For open-shell nuclei, the BCS method is used to deal with the pairing correlation between nucleons, only considering the same nucleon pairing ($ nn $ or $ pp $), using the Gogny interaction D1S \cite{J.F.Berger-NPA428(1984)23, Bender.M-EPJA8(2000)59}. Additionally, for hypernuclei with an odd number of nucleons, the blocking effect should be considered for the last valence nucleon or $ \Lambda_{c}^{+} $ hyperon, as detailed in Ref. \cite{Qian.Zhuang-PRC106(2022)054311}. In this work, the $ \Lambda_{c}^{+} $ hyperon occupy the $ 1s_{1/2} $ orbital.

\subsection{$ \Lambda_cN $ effective interaction and stability of charmed hypernuclei}

\begin{figure}
    \centering
    \includegraphics[width=0.48\textwidth]{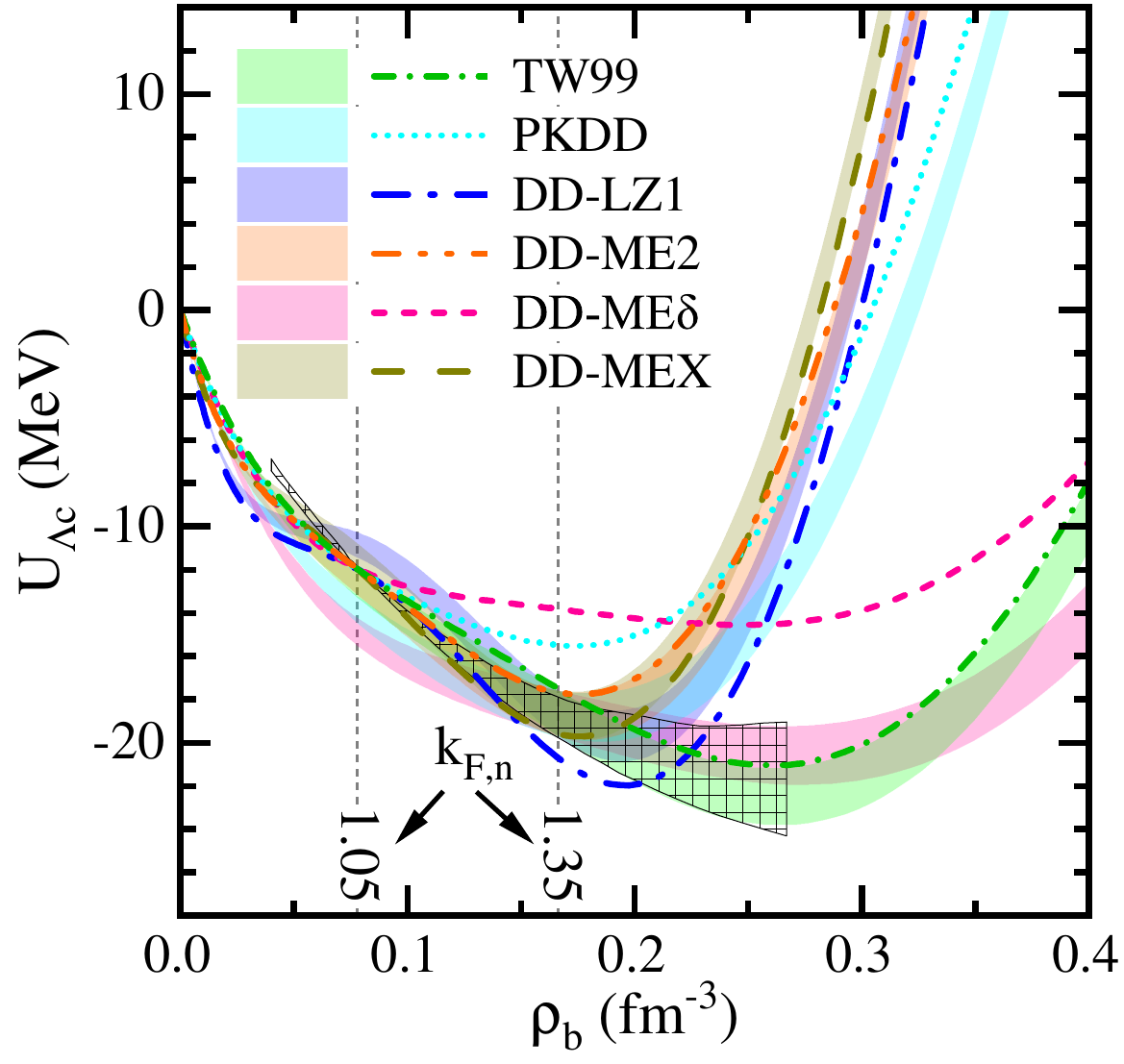}
    \caption{The $\Lambda_c^+$ potentials ${U}_{\Lambda_{c}}$ as a function of the baryon density $\rho_b$ in symmetric nuclear matter, calculated by the $\Lambda_{c}N$ effective interactions within DDRMF models. The lines represent the results from $\Lambda_{c}1$, while the shaded regions correspond to $\Lambda_{c}2$. The square grid area indicates the empirical constraints of the $\Lambda_c^+$ potential extracted from Fig.1(a) in Ref. \cite{Haidenbauer.J-EPJA56(2020)195}.}
    \label{Fig:Fig1-Pot_in_NM}
\end{figure}

To achieve a theoretical description of hypernuclear structure, further development of the $ \Lambda_{c} N $ interaction is required within the framework of RMF theory, this interaction can generally be expressed as the ratio of coupling strengths between meson-hyperon and meson-nucleon. Since $ \Lambda_{c}^{+} $ is an isospin-zero particle, only isoscalar mesons can participate in the $ \Lambda_{c} N $ interaction within the meson-exchange diagram. Specifically, for the isoscalar vector $ \omega^{\mu} $ meson, the ratio of coupling strength is determined to be $ R_{\omega \Lambda_{c}} = g_{\omega \Lambda_{c}}/g_{\omega N} = 0.666 $ based on the n\"{a}ive quark model \cite{Tan.Yu.Hong-PRC70(2004)054306, Yu.Hong.Tan-Europhys.Lett67(2004)355}. As for the isoscalar scalar $ \sigma $ meson, its coupling strength is generally determined from experimental data. However, the available experimental data for the $ \Lambda_{c}^{+} $ hypernucleus is currently quite limited, and there is insufficient evidence to confirm the observation of bound $ \Lambda_{c}^{+} $ hypernuclei. Thus, alternative approaches are needed to construct the $ \Lambda_{c} N $ interaction. As mentioned in the introduction, results obtained from lattice QCD simulations combined with chiral effective field theory extrapolation provide a reasonable reference \cite{Haidenbauer.J-EPJA54(2018)199, Haidenbauer.J-EPJA56(2020)195}. To investigate the properties of the $ \Lambda_{c} N $ interaction in nuclear matter, subsequent work has used the Brueckner-Hartree-Fock method to obtain the $ \Lambda_{c}^{+} $ hyperon potential $ U_{\Lambda_{c}} $ in symmetric nuclear matter at finite density \cite{Haidenbauer.J-EPJA56(2020)195}. This provides an excellent bridge for studying $ \Lambda_{c}^{+} $ hypernuclei based on first-principle calculations and RMF theory. By fitting the empirical $ \Lambda_{c}^{+} $ hyperon potential obtained from first-principles calculations at specific densities or Fermi momenta, the $ \Lambda_{c} N $ interaction can be effectively constructed within the framework of DDRMF theory. Then, the $ \Lambda_{c}^{+} $ hyperon potential in symmetric nuclear matter can be expressed as
\begin{align}\label{eq:pot_in_NM}
	U_{\Lambda_{c}} & =\sum_{B}\left[  -g_{\sigma\Lambda_c}\frac{g_{\sigma B}}{m_{\sigma}^2}\rho_{s,B} + g_{\omega\Lambda_c}\frac{g_{\omega B}}{m_{\omega}^2}\rho_{b,B} \right.\nonumber\\
	& +\frac{1}{\rho_{0}}\left.\left( -\frac{g_{\sigma B}}{m_{\sigma}^2}\rho_{s,B}^2\frac{\partial g_{\sigma B}}{\partial \xi} + \frac{g_{\omega B}}{m_{\omega}^2}\rho_{b,B}^2\frac{\partial g_{\omega B}}{\partial \xi}\right) \right].
\end{align}
Then, the ratio of the $ \sigma$-$\Lambda_{c} $ coupling strength can be determined by fitting the empirical hyperon potential in symmetric nuclear matter, as presented in Fig. 1(a) of Ref. \cite{Haidenbauer.J-EPJA56(2020)195}. It is worth noting that due to the utilization of different cutoff values ($ \Lambda = 500 $ MeV or $ \Lambda = 600 $ MeV) in handling the hyperon-nucleon interaction, there exists some uncertainty in the evolution of the empirical hyperon potential with Fermi momentum. For clarity in discussion, data from the empirical hyperon potential has been extracted and showcased as a square grid area in Fig. \ref{Fig:Fig1-Pot_in_NM}. Notably, at a Fermi momentum of $ k_{F,n} = 1.05~\mathrm{fm}^{-1} $, the empirical hyperon potentials coincide for both cutoffs, while differences between the potentials obtained with different cutoffs become more pronounced as the Fermi momentum moves towards smaller or larger regions. To minimize the introduction of additional influences when constructing the $ \Lambda_{c} N $ interaction, we first fit the empirical hyperon potential at Fermi momentum $ {k_{F,n}} = 1.05~\mathrm{fm}^{-1} $ with $ U_{\Lambda_{c}} = -11.98 $ MeV, resulting in a series of effective $ \Lambda_{c} N $ interactions labeled as $ \Lambda_{c}1 $. Furthermore, considering that the effective $ \Lambda_{c} N $ interactions will ultimately be applied in describing hypernuclear structures, where the nuclear medium density is comparable to the saturation density, it suggests that the effective $ \Lambda_{c} N $ interaction obtained by fitting the empirical hyperon potential at saturation density might provide a more reliable description for hypernuclear structures. Therefore, we additionally selected the empirical hyperon potential at the Fermi momentum corresponding to saturation density, $ {k_{F,n}} = 1.35~\mathrm{fm}^{-1} $, as a new fitting target, resulting in the effective $ \Lambda_{c} N $ interaction denoted as $ \Lambda_{c}2 $. Due to uncertainties in the empirical hyperon potential arising from different cutoffs, $ \Lambda_{c}2 $ exhibits a certain degree of uncertainty. Its upper and lower limits are obtained by fitting the upper limit $ U_{\Lambda_{c}} = -19.70 $ MeV and the lower limit $ U_{\Lambda_{c}} = -17.60 $ MeV of the empirical hyperon potential, respectively.

Based on the selection of several DDRMF effective interactions, two different fitting strategies were employed to determine the ratio of the $ \sigma $-$ \Lambda_{c} $ coupling strength $ R_{\sigma\Lambda_{c}} $, as shown in Table \ref{Coupling strengths}. Additionally, the table summarizes the effective masses $ {M}^{\ast}_{\Lambda_{c}}/{M}_{\Lambda_{c}} $ for hyperons obtained from various effective interactions, as well as the hyperon potential $ U_{\Lambda_{c}} $ at saturation density. It can be observed that for both $ \Lambda_{c}1 $ and $ \Lambda_{c}2 $, the $ R_{\sigma\Lambda_{c}} $ provided by various DDRMF effective interactions exhibit significant differences, which affect the description of nuclear matter properties such as the effective mass and hyperon potential. Due to the relatively large mass of the $ \Lambda_{c}^{+} $ particle, the effective mass, despite showing some differences, does not exhibit a sensitive dependence on $ R_{\sigma\Lambda_{c}} $. In contrast, the hyperon potential is significantly dependent on the strength of the $ \Lambda_{c} N $ interaction, especially for the effective interaction $ \Lambda_{c}1 $. Although all the DDRMF Lagrangians are fitted to the empirical hyperon potential at the minimum uncertainty, there are significant variations in the hyperon potential at saturation density as the baryon density evolves, ranging from $ -13.65 $ MeV to $ -19.98 $ MeV for different effective interactions. To reduce theoretical uncertainties, it is necessary to identify some model-sensitive observables in the future to impose additional constraints.

\begin{table}[hbpt]
    \centering
    \caption{The $ \sigma$-$\Lambda_{c} $ coupling strengths $ {R}_{\sigma\Lambda_{c}} $ fitted for several DDRMF effective interactions according to the empirical constraints of the $ \Lambda_c^+ $ potential $ {U}_{\Lambda_{c}} $ in symmetric nuclear matter \cite{Haidenbauer.J-EPJA56(2020)195}. In detail, the series $ \Lambda_{c}1 $ is determined by the fixed potential $ {U}_{\Lambda_{c}}=-11.98 $ MeV at $ k_{F,n}=1.05~\rm{fm}^{-1} $, while two values of $ \Lambda_{c}2 $ which define the lower and upper limits are given by fitting $ {U}_{\Lambda_{c}}=-17.60 $ or $ -19.70 $ MeV at $ k_{F,n}=1.35~\rm{fm}^{-1} $, respectively. In addition, the $ \Lambda_c^+ $ Dirac effective masses $ {M}^{\ast}_{\Lambda_{c}}/{M}_{\Lambda_c} $ and $ \Lambda_c^+ $ potentials $ {U}_{\Lambda_{c}} $ (in MeV) in symmetric nuclear matter at saturation density are summarized as well.}\label{Coupling strengths}
    \renewcommand{\arraystretch}{1.5}
    \setlength{\tabcolsep}{6pt}
    \begin{tabular}{ccccc}
\hline\hline
        &                                          &$\Lambda_{c}1$           & \multicolumn{2}{c}{$\Lambda_{c}2$} \\\hline
\multirow{3}{*}{TW99}
        &${R}_{\sigma\Lambda_{c}}$                 & 0.5847                  & 0.5849               & 0.5897\\
        &${M}^{\ast}_{\Lambda_{c}}/{M}_{\Lambda_c}$& 0.8936                  & 0.8936               & 0.8927\\
        &${U}_{\Lambda_{c}}$                       & -16.69                  & -16.77               & -18.78\\\hline
\multirow{3}{*}{PKDD}
        &${R}_{\sigma\Lambda_{c}}$                 & 0.5836                  & 0.5885               & 0.5934\\
        &${M}^{\ast}_{\Lambda_{c}}/{M}_{\Lambda_c}$& 0.8977                  & 0.8968               & 0.8960\\
        &${U}_{\Lambda_{c}}$                       & -15.21                  & -17.18               & -19.16\\\hline
\multirow{3}{*}{DD-LZ1}
        &${R}_{\sigma\Lambda_{c}}$                 & 0.5908                  & 0.5836               & 0.5885\\
        &${M}^{\ast}_{\Lambda_{c}}/{M}_{\Lambda_c}$& 0.8924                  & 0.8937               & 0.8928\\
        &${U}_{\Lambda_{c}}$                       & -19.98                  & -17.00               & -19.03\\\hline
\multirow{3}{*}{DD-ME2}
        &${R}_{\sigma\Lambda_{c}}$                 & 0.5878                  & 0.5876               & 0.5925\\
        &${M}^{\ast}_{\Lambda_{c}}/{M}_{\Lambda_c}$& 0.8968                  & 0.8968               & 0.8959\\
        &${U}_{\Lambda_{c}}$                       & -17.24                  & -17.16               & -19.13\\\hline
\multirow{3}{*}{DD-MEX}
        &${R}_{\sigma\Lambda_{c}}$                 & 0.5902                  & 0.5857               & 0.5905\\
        &${M}^{\ast}_{\Lambda_{c}}/{M}_{\Lambda_c}$& 0.8923                  & 0.8931               & 0.8922\\
        &${U}_{\Lambda_{c}}$                       & -18.98                  & -17.10               & -19.11\\\hline
\multirow{3}{*}{DD-ME$\delta$}
        &${R}_{\sigma\Lambda_{c}}$            & 0.5799                  & 0.5896               & 0.5951\\
        &${M}^{\ast}_{\Lambda_{c}}/{M}_{\Lambda_c}$& 0.9069                  & 0.9053               & 0.9044\\
        &${U}_{\Lambda_{c}}$                       & -13.65                  & -17.21               &-19.23\\\hline\hline
\end{tabular}
\end{table}

To gain a more intuitive understanding of the differences in the effective $ \Lambda_{c} N $ interactions among various DDRMF models, Fig. \ref{Fig:Fig1-Pot_in_NM} shows the evolution of the $ \Lambda_{c}^{+} $ hyperon potential as a function of baryon density $ \rho_{b} $ in symmetric nuclear matter. The lines represent the results for effective interaction $ \Lambda_{c}1 $, while the results for $ \Lambda_{c}2 $ are indicated by the shaded region. For $\Lambda_{c}1$, the interaction is obtained by fitting the empirical hyperon potential at a Fermi momentum of $ k_{F,n} = 1.05~\rm{fm}^{-1} $, resulting in a general consistency of the hyperon potential at low densities across various DDRMF models, except for DD-LZ1. However, as the hyperon potential evolves towards higher baryon densities, discrepancies among the models quickly emerge, with significant differences already evident in the subsaturation region. For $ \Lambda_{c}2 $, the interaction is derived by fitting the empirical hyperon potential at $ k_{F,n} = 1.35~\rm{fm}^{-1} $, and its results show significant model dependence both at high and low densities. Since this work focuses on the existence of $ \Lambda_{c}^{+} $ hypernuclei and their structural properties, we primarily examine the behavior of the $ \Lambda_{c} N $ effective interaction in regions below saturation density. By analyzing the results of $ \Lambda_c1 $ and $ \Lambda_c2 $, we can understand the impact of the uncertainty in the hyperon potential at saturation and low densities on the description of hypernuclear structures. It is worth noting that both $ \Lambda_c1 $ and $ \Lambda_c2 $ show the most significant differences in the hyperon potential below saturation density for the DD-LZ1 and DD-ME$ \delta $ models, which might reflect the uncertainty of the DDRMF theory in regions below saturation density. In contrast, TW99 and DD-ME2 exhibit the smallest differences in the hyperon potential below saturation density, and the results for $ \Lambda_c1 $ and $ \Lambda_c2 $ are largely consistent, making these models ideal for a self-consistent and unified description of charmed nuclear matter properties and charmed hypernuclear structures below saturation density. Therefore, in the subsequent discussion, we will select these four typical effective interactions for in-depth analysis. Additionally, considering the potential application to the properties of charmed nuclear matter, a brief summary of the hyperon potential in regions above saturation density is provided. It shows that the evolution of the hyperon potential with baryon density obtained from various DDRMF functionals primarily exhibits two forms. In some cases, the $ \Lambda_{c}^{+} $ hyperon potential transitions from attractive to repulsive with increasing baryon density and rises rapidly, while in other cases, the hyperon potential evolves slowly with baryon density and remains weakly attractive at higher densities.

\begin{figure}
    \centering
    \includegraphics[width=0.48\textwidth]{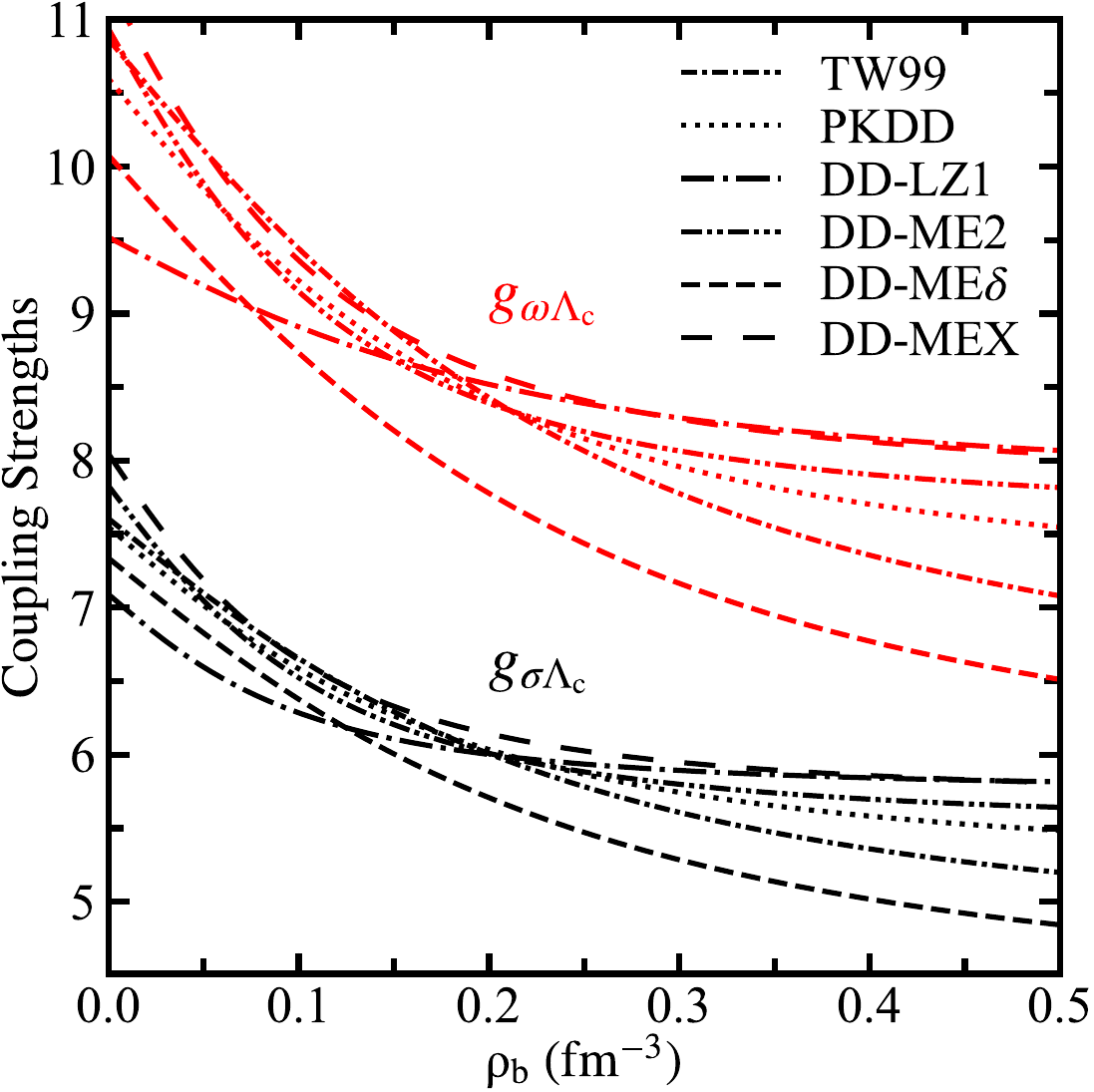}
	\caption{The baryon density dependence of meson-$\Lambda_{c}$ coupling strengths, namely, the isoscalar $g_{\sigma\Lambda_{c}}$ (black lines) and  $g_{\omega\Lambda_{c}}$ (red lines), for the $\Lambda_{c}N$ effective interactions $\Lambda_c$1 within DDRMF models.}
    \label{Fig:Fig2-Coupling_constant}
\end{figure}

The significant differences in the hyperon potential in symmetric nuclear matter can be qualitatively explained by the density-dependent behavior of the coupling constants. As shown in Fig. \ref{Fig:Fig2-Coupling_constant}, taking $ \Lambda_{c}1 $ as an example, the meson-hyperon coupling strengths varying with baryon density for different effective interactions are presented. According to Eq. \eqref{eq:pot_in_NM}, it shows that the hyperon potential depth depends not only on the magnitude of the coupling strength but also on its rate of change. To clearly reflect the influence of these two factors on the potential, we select TW99, PKDD and DD-ME$ \delta $ for analysis. For TW99 and PKDD which have the same value of the coupling strength except the slope around $ \rho_{b} = 0.2~\rm{fm^{-3}} $, they can be used to explain the potential discrepancy caused by the rate of change. At the same time, TW99 and DD-ME$ \delta $ are selected to illustrate the effect of magnitude on the hyperon potential given their similar rate of change. Firstly, we will discuss the effect of changes in coupling strengths on the hyperon potential. From Fig. \ref{Fig:Fig2-Coupling_constant}, it shows that the variation of the coupling strength for DD-ME2 is larger than PKDD, indicating that DD-ME2 has a large potential depth. In fact, the magnitude of $ g_{\sigma N} $ also has an influence on the hyperon potential, because these two models have different $ R_{\sigma\Lambda} $ with the value for TW99 being grater than that for PKDD as shown in Table \ref{Coupling strengths}. The smaller $ R_{\sigma\Lambda} $ means that PKDD has a larger $ g_{\sigma N} $, which can lead to extra repulsion in the hyperon potential. To study the effect of the magnitude in coupling strength on the potential depth, attention is focused on the two models TW99 and DD-ME$ \delta $. It can be seen in Fig. \ref{Fig:Fig2-Coupling_constant} that the value of coupling strengths for TW99 is larger than DD-ME$ \delta $, and the large coupling strengths indicate TW99 has a deeper potential. Additionally, following the above analysis, smaller $ R_{\sigma\Lambda_c} $ given by DD-ME$ \delta $ will further results in shallower potential depth compared with TW99. Finally, it can be seen that the coupling strengths of TW99 and DD-ME$ \delta $ decrease more rapidly than other effective interactions. This tendency makes the two models still have an attractive potential at higher density.

\begin{figure}
    \centering
    \includegraphics[width=0.48\textwidth]{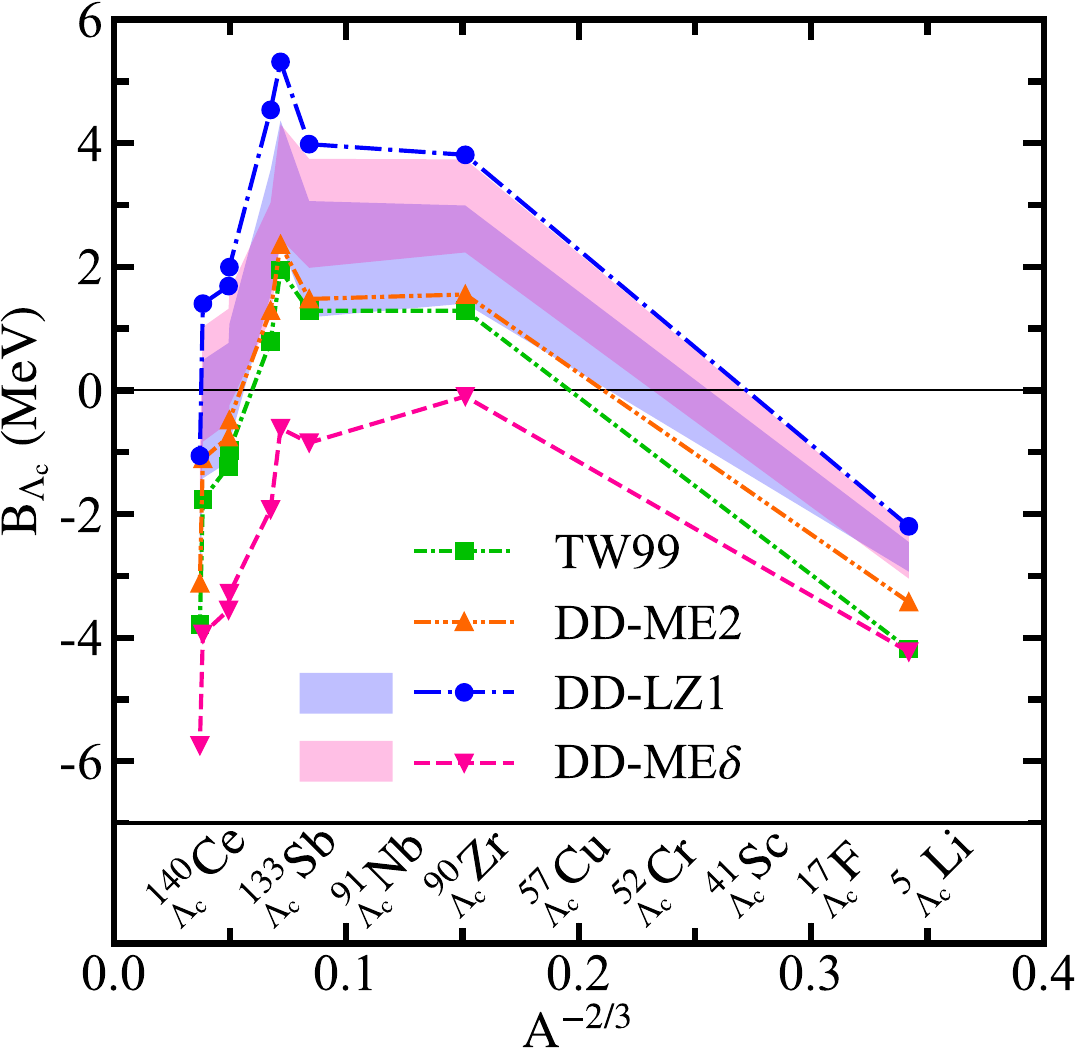}
    \caption{The calculated $\Lambda_c^+$ separation energies $B_{\Lambda_{c}}$ for the ground state of charmed hypernuclei with various DDRMF effective interactions. The lines represent the results of the $\Lambda_c$1 model, whereas the shaded areas with just two examples (DD-LZ1 and DD-ME$\delta$) correspond to those from $\Lambda_c$2.}
    \label{Fig:Fig3-SeparationEnergy}
\end{figure}

\begin{figure}
    \centering
    \includegraphics[width=0.48\textwidth]{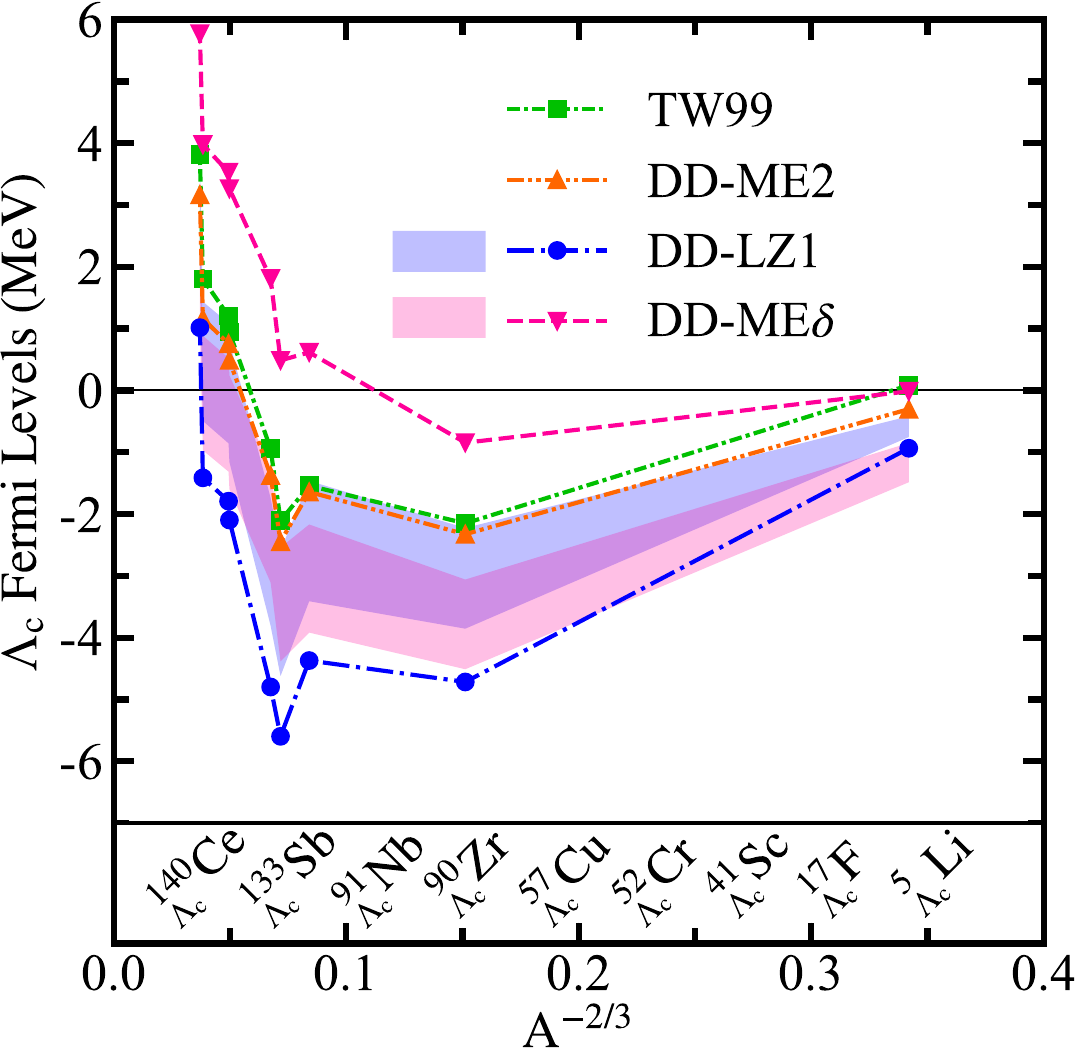}
    \caption{Similar to Fig. \ref{Fig:Fig3-SeparationEnergy} but for $\Lambda_c^+$ Fermi energies.}
    \label{Fig:Fig4-FermiLevels}
\end{figure}

To accurately describe hypernuclear structures, we will first conduct an in-depth discussion on the existence of bound $ \Lambda_{c}^{+} $ hypernuclei using the four typical DDRMF Lagrangians, namely TW99, DD-ME2, DD-LZ1, and DD-ME$ \delta $, combined with the $ \Lambda_{c} N $ effective interaction. Generally, the existence of bound hypernuclei can be determined by the separation energy $ B_{\Lambda_{c}} $ and the Fermi level of the hyperon. The separation energy $ B_{\Lambda_{c}} $ is defined as the difference in binding energies, expressed as follows

\begin{equation}\label{eq:binding-energy}
	B_{\Lambda_{c}}\left[ ^{A}_{\Lambda_{c}}Z \right] \equiv E\left[ ^{A-1}{(Z-1)} \right] - E\left[ ^{A}_{\Lambda_{c}}Z \right],
\end{equation}
Among them, the nucleonic core is represented as $ ^{A-1}{(Z-1)} $, and the corresponding hypernucleus as $ ^A_{\Lambda_{c}}Z $. Based on the aforementioned four sets of DDRMF Lagrangians and the $ \Lambda_{c} N $ effective interaction listed in Table \ref{Coupling strengths}, the hyperon separation energies and Fermi levels for charmed hypernuclei with different mass numbers are presented in Fig. \ref{Fig:Fig3-SeparationEnergy} and Fig. \ref{Fig:Fig4-FermiLevels}, respectively. Among the various DDRMF functionals, the most significant differences in the hyperon potentials are observed between DD-LZ1 and DD-ME$ \delta $, which greatly influence the description of the $ \Lambda_{c}^{+} $ hypernuclear bulk and single-particle properties, and these differences can be considered as sources of uncertainty in the DDRMF theory. Therefore, only the results based on DD-LZ1 and DD-ME$ \delta $ are shown for $ \Lambda_{c}2 $. From both Fig. \ref{Fig:Fig3-SeparationEnergy} and Fig. \ref{Fig:Fig4-FermiLevels}, it is evident that the results for $ \Lambda_{c}1 $, corresponding to different DDRMF Lagrangians, exhibit stronger model dependence compared to $ \Lambda_{c}2 $. This is because $ \Lambda_{c}1 $ is obtained by fitting the empirical hyperon potential at $ k_{F,n} = 1.05~\mathrm{fm}^{-1} $, corresponding to a relatively low baryon density. As the baryon density evolves towards the region of saturation density, various DDRMF functionals quickly diverge in their predictions for the hyperon potential. In fact, the hyperons in bound $ \Lambda_{c}^{+} $ hypernuclei are mainly distributed within the nucleus, where the nuclear medium density is close to saturation density. Consequently, the description of hyperon separation energies and Fermi levels is closely related to the effective $ \Lambda_{c} N $ interaction at saturation density. Comparing the results of several DDRMF Lagrangians, it can be seen that for hyperon separation energies and Fermi levels, TW99 and DD-ME2 yield very similar results. This consistency is due to the similar behavior of their hyperon potentials under symmetric nuclear matter. However, DD-LZ1 and DD-ME$ \delta $ show significant differences in their results. Compared to $ \Lambda_{c}1 $, the effective interaction $ \Lambda_{c}2 $ obtained by fitting the empirical hyperon potential near saturation density significantly reduces uncertainty, indicating that it may introduce less model dependence in the description of hypernuclear structures. Furthermore, the maximum hypernuclear separation energy obtained from the DDRMF functionals is approximately $ 5 $ MeV, which closely aligns with the conclusion of $ 0.6 U_{\Lambda} $ in Ref. \cite{Liu.Yi.Xiu-PRC108(2023)064312}.

Based on the selected DDRMF models and the developed $\Lambda_{c} N$ effective interaction, a systematic evaluation of the existence of charmed hypernuclei was conducted. The criterion for evaluation is the $\Lambda_{c}^{+}$ separation energy $B_{\Lambda_{c}} > 0$, with the results presented in Table \ref{Tab:Lc_Stability}. Specifically, the symbols “$+$” and “$-$” denote the presence and absence of $\Lambda_{c}^{+}$ hypernuclei, respectively. Due to the uncertainty in $\Lambda_{c}2$, some models predict bound charmed hypernuclei only under certain conditions, indicated by the symbol “$\bigcirc$”. For $\Lambda_{c}1$, the evaluation results show significant model dependence. For example, the DD-ME$\delta$-$\Lambda_{c}1$ model, which has the shallowest hyperon potential in nuclear matter near saturation density, predicts that $\Lambda_{c}^{+}$ hypernuclei are unbound. Conversely, DD-LZ1-$ \Lambda_{c}1 $ model predicts the broadest range of hypernuclear existence, extending from $ ^{17}_{\Lambda_{c}} $F to $ ^{133}_{\Lambda_{c}} $Sb, due to its relatively deeper potential. Regarding $ \Lambda_{c}2 $, it is evident that all DDRMF models produce the same results, suggesting that the $ \Lambda_{c}^{+} $ hyperon resides in a high-density region inside the nucleus. Additionally, the range of hypernuclear existence for $ \Lambda_{c}2 $ is predominantly influenced by uncertainties in the potential depth. When using $ {U}_{\Lambda_{c}} = -19.70 $ MeV to determine the $ \Lambda_{c} N $ interaction, the heaviest possible hypernucleus is $ ^{133}_{\Lambda_{c}}\mathrm{Sb} $. However, with a shallower potential of $ {U}_{\Lambda_{c}} = -17.60 $ MeV, the heaviest hypernucleus is limited to $ ^{57}_{\Lambda_{c}}\mathrm{Cu} $.

\begin{table}[hbpt]
    \centering
    \caption{The summary for the existence of various charmed hypernuclei with DDRMF models (coupled for both $\Lambda_c$1 and $\Lambda_c$2), which is evaluated by the criterion that $\Lambda_c^+$ separation energies $B_{\Lambda_{c}}>0$. As a rule, fulfillment (violation) of the constraint is indicated with ``$+(-)$" and the marginal cover just in the case of $\Lambda_c$2 is marked with ``$\bigcirc$". See the text for details.}\label{Tab:Lc_Stability}
    \renewcommand{\arraystretch}{1.50}
    \setlength{\tabcolsep}{1pt}
    \begin{tabular}{ccccccc}
        \hline\hline
                                       &TW99&PKDD&DD-LZ1&DD-ME$\delta$&DD-ME2&DD-MEX\\\hline
        $^{5}_{\Lambda_c}\mathrm{Li}   $&$-$$-$&$-$$-$       &$-$$-$     &$-$$-$       &$-$$-$       &$-$$-$     \\
        $^{17}_{\Lambda_c}\mathrm{F}   $&++    &++           &++         &$-$+         &++           &++         \\
        $^{41}_{\Lambda_c}\mathrm{Sc}  $&++    &++           &++         &$-$+         &++           &++         \\
        $^{52}_{\Lambda_c}\mathrm{Cr}  $&++    &++           &++         &$-$+         &++           &++         \\
        $^{57}_{\Lambda_c}\mathrm{Cu}  $&++    &$-$+         &++         &$-$+         &++           &++         \\
        $^{90}_{\Lambda_c}\mathrm{Zr}
        $&$-$$\bigcirc$&$-$$\bigcirc$&+$\bigcirc$&$-$$\bigcirc$&$-$$\bigcirc$&+$\bigcirc$\\
        $^{91}_{\Lambda_c}\mathrm{Nb} $&$-$$\bigcirc$&$-$$\bigcirc$&+$\bigcirc$&$-$$\bigcirc$&$-$$\bigcirc$&+$\bigcirc$\\
        $^{133}_{\Lambda_c}\mathrm{Sb}$&$-$$\bigcirc$&$-$$\bigcirc$&+$\bigcirc$&$-$$\bigcirc$&$-$$\bigcirc$&+$\bigcirc$\\
        $^{140}_{\Lambda_c}\mathrm{Ce}$&$-$$-$       &$-$$-$       &$-$$-$     &$-$$-$       &$-$$-$       &$-$$-$     \\ \hline\hline
    \end{tabular}
\end{table}

\subsection{Properties of the charmed hypernuclei}

\begin{figure}
    \centering
    \includegraphics[width=0.48\textwidth]{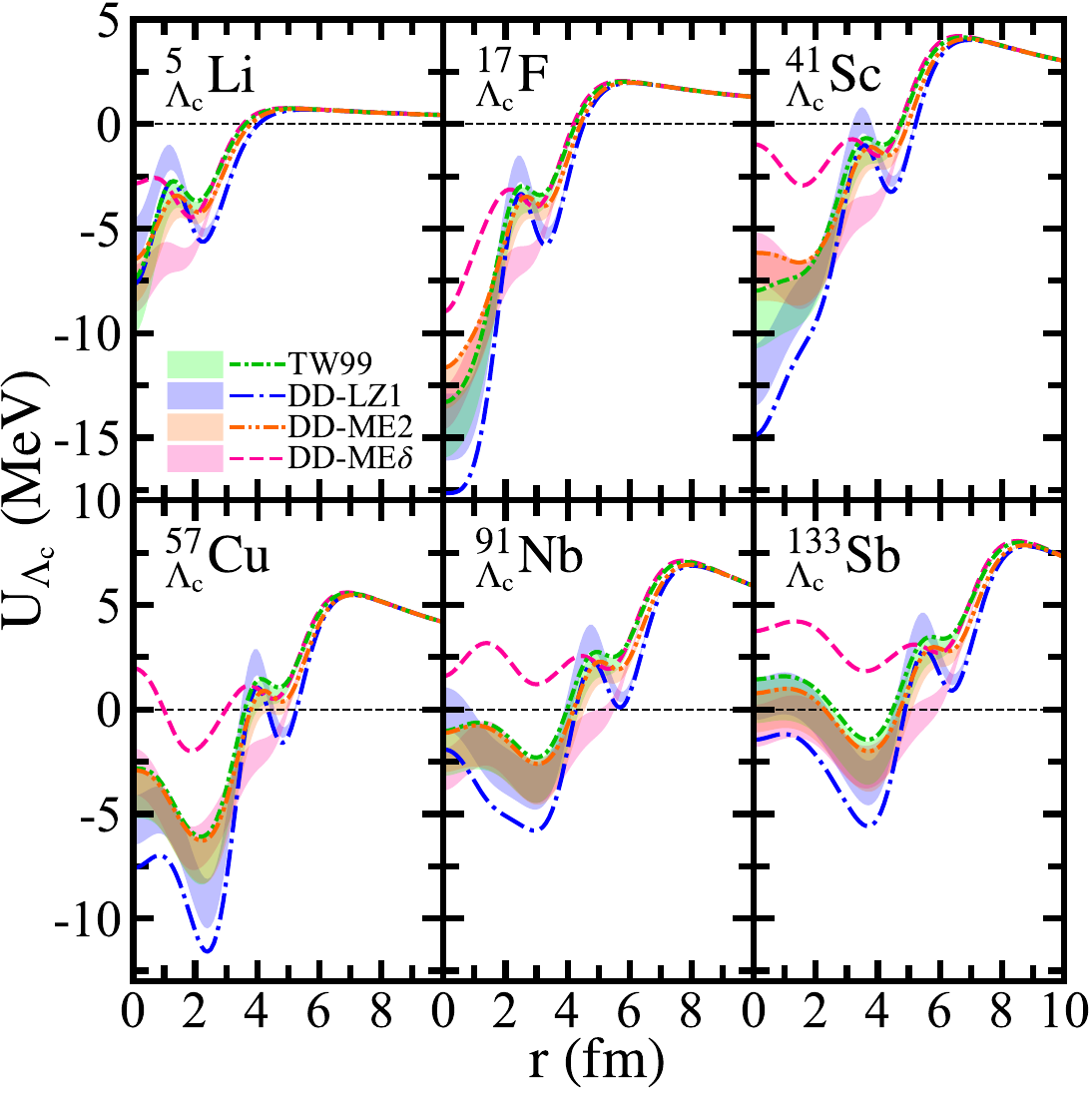}
    \caption{The $\Lambda_c^+$ mean-field potentials in charmed hypernuclei $^5_{\Lambda_c}$Li, $^{17}_{\Lambda_c}$F, $^{41}_{\Lambda_c}$Sc, $^{57}_{\Lambda_c}$Cu, $^{91}_{\Lambda_c}$Nb, $^{133}_{\Lambda_c}$Sb as a function of radial coordinate $r$ with various DDRMF effective interactions. The lines represent the results of the $\Lambda_c$1 model, whereas the shaded areas correspond to those from $\Lambda_c$2.}
    \label{Fig5:Lc_Potentials}
\end{figure}

To further understand the bulk and single-particle properties of hypernuclei, based on four typical DDRMF models, namely TW99, DD-ME2, DD-LZ1, and DD-ME$\delta$, the hyperon potentials for several representative $\Lambda_{c}^{+}$ hypernuclei are presented, as shown in Fig. \ref{Fig5:Lc_Potentials}. It can be seen that as the mass number increases, the depth of the hyperon potentials initially increases and then decreases, even resulting in unbound hyperon potentials. Comparing the various DDRMF models, it is found that the DD-ME$\delta$-$\Lambda_{c}1$ model provides a relatively shallow hyperon potential, whereas DD-LZ1-$\Lambda_{c}1$ exhibits the opposite trend. As a result, DD-ME$\delta$-$\Lambda_{c}1$ predicts unbound results for the studied $\Lambda_{c}^{+}$ hypernuclei, while DD-LZ1-$\Lambda_{c}1$ suggests the existence of $\Lambda_{c}^{+}$ hypernuclei across a wider range of masses. Furthermore, it is observed across all DDRMF models that the hyperon potentials generally exhibit peaks near the surface. These peaks extend to larger radial distances with increasing hypernuclear mass number, and at these larger radial distances, the hyperon potentials consistently contribute repulsively. Further analysis of this phenomenon will be conducted in subsequent discussions.

\begin{figure}
    \centering
    \includegraphics[width=0.48\textwidth]{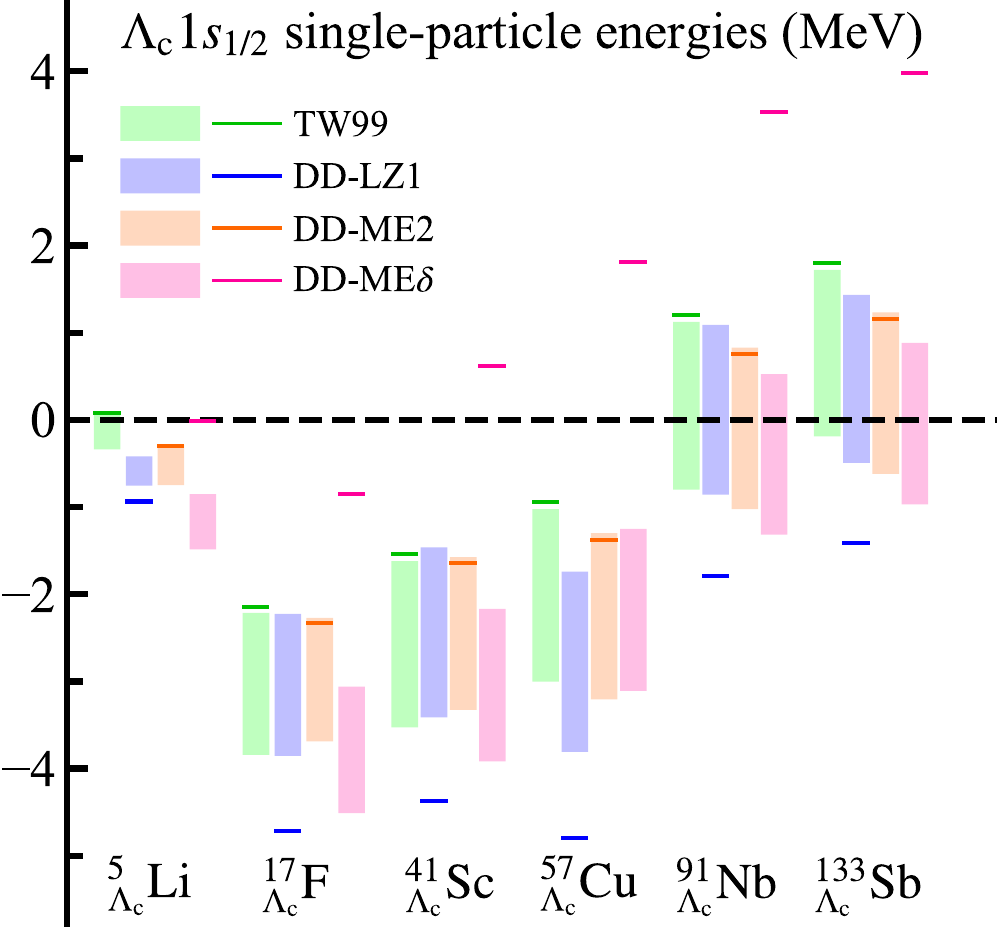}
    \caption{The $\Lambda_c^+$ single-particle energies of $1s_{1/2}$ state in charmed hypernuclei $^5_{\Lambda_c}$Li, $^{17}_{\Lambda_c}$F, $^{41}_{\Lambda_c}$Sc, $^{57}_{\Lambda_c}$Cu, $^{91}_{\Lambda_c}$Nb, $^{133}_{\Lambda_c}$Sb are presented with the selected DDRMF effective interactions. The horizontal lines depict the results of $\Lambda_{c}1$, and the histogram areas represent ones from $\Lambda_{c}2$.}
    \label{Fig6-LcSingle_particle_energy.pdf}
\end{figure}

In conjunction with the hyperon potentials, the corresponding $ \Lambda_{c}^{+} $ single-particle energies for the $ 1s_{1/2} $ state are depicted in Fig. \ref{Fig6-LcSingle_particle_energy.pdf}. Firstly, it is observed that the single-particle energies initially decrease and then increase, aligning with the trend of the hyperon potentials shown in Fig. \ref{Fig5:Lc_Potentials}. Furthermore, for $ \Lambda_{c}1 $, the energy levels given by different models show significant variations compared to those for $ \Lambda_{c}2 $. This results also indicate that the uncertainty in hyperon potentials within nuclear matter at $ k_{F,n} = 1.35~\rm{fm}^{-1} $, as illustrated in Fig. \ref{Fig:Fig1-Pot_in_NM}, significantly influences the single-particle properties of the $ \Lambda_{c}^{+} $ hyperon. In addition to the $ \Lambda_{c} N $ interaction, differences in $ NN $ interactions may also significantly affect the stability of the $ \Lambda_{c}^+ $ hypernucleus. When the $ \Lambda_{c} N $ interaction is specified, that is, when fitting the same $ U_{\Lambda_{c}} $, we observe that the $ NN $ interactions from these models exhibit varying density-dependent behaviors. Variations in the $ NN $ interactions could potentially alter the distribution of the $ \Lambda_{c}^{+} $ potential, which in turn could affect the single-particle properties of the hyperon. At the same time, we have chosen effective interactions that are more consistent with the range of $ \Lambda_{c}^{+} $ potentials reported by \cite{Haidenbauer.J-EPJA56(2020)195} as a comparison, specifically TW99 and DD-ME2. We find the trend in the variation of single-particle energy levels is consistent with the literature; however, our models indicate weak binding for the hyperon.

\begin{figure}
    \centering
    \includegraphics[width=0.48\textwidth]{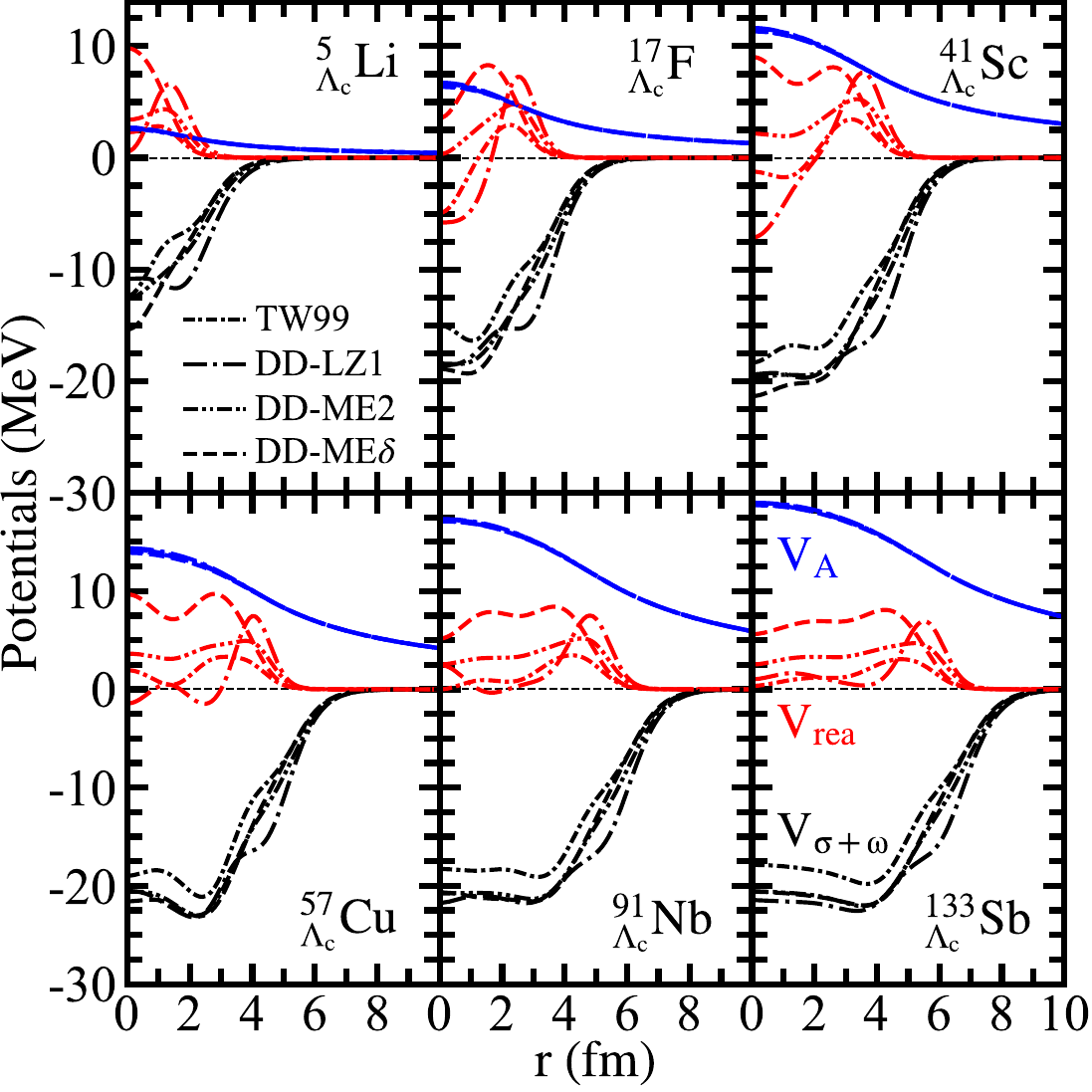}
    \caption{The decomposition of the $\Lambda_c^+$ mean-field potentials for charmed hypernuclei $^5_{\Lambda_c}$Li, $^{17}_{\Lambda_c}$F, $^{41}_{\Lambda_c}$Sc, $^{57}_{\Lambda_c}$Cu, $^{91}_{\Lambda_c}$Nb, $^{133}_{\Lambda_c}$Sb, using four selected DDRMF effective interactions. The contributions from $\sigma$ and $\omega$ mesons are denoted as $V_{\sigma+\omega}$ (black curves), the Coulomb potentials as $V_\mathrm{A}$ (blue), and those from the rearrangement terms as $V_{\mathrm{rea}}$ (red) due to the density dependence of meson-$\Lambda_{c}$ couplings.}
    \label{Fig7-decomposition_potential.pdf}
\end{figure}

To clarify the factors affecting hypernuclear stability, the contributions to the hyperon potential can be further decomposed. Since $\Lambda_{c}^{+}$ is a positively charged particle with zero isospin, the hyperon potential includes contributions from isoscalar mesons ($ \sigma $ and $ \omega^{\mu} $) and photon. Additionally, the rearrangement term arising from the density dependence of the meson-baryon coupling strengths need to be considered. Based on four typical DDRMF models, and using $\Lambda_{c}1$ as an example, a series of hypernuclei ranging from $^{5}_{\Lambda_c}$Li to $^{133}_{\Lambda_{c}}$Sb were selected. The hyperon potential was decomposed into contributions from isoscalar mesons ($ V_{\sigma+\omega} $), Coulomb interaction ($ V_{\rm{A}} $), and rearrangement terms ($ V_{\rm{rea}} $), as shown in Fig. \ref{Fig7-decomposition_potential.pdf}. It can be observed that as the mass number of the hypernuclei increases, $V_{\sigma+\omega}$ gradually deepens and saturates at around $ -20 $ MeV. The contribution from Coulomb repulsion $ V_{\rm{A}} $ also gradually increases, becoming a significant factor affecting the stability of heavy hypernuclei. As for the rearrangement terms $ V_{\rm{rea}} $, they mainly provide repulsive contributions, peaking at about $ 10 $ MeV near the hypernuclear surface. This phenomenon accounts for the presence of peaks in the hyperon potential near the nuclear surface. Comparing the contributions in light hypernuclei, it is found that the repulsive contribution from the rearrangement term is comparable to, or even exceeds, the Coulomb interaction. This indicates the crucial role of the rearrangement term in describing the stability of light hypernuclei. In fact, the contribution of the rearrangement term can be qualitatively explained by the density-dependent behavior of the coupling strengths shown in Fig. \ref{Fig:Fig2-Coupling_constant} and Eq. \eqref{eq:Erea}. At the nuclear surface, the drastic evolution of the coupling strength in low-density situations significantly alters the contribution of the rearrangement term. Besides influencing the stability of hypernuclei, the rearrangement term also plays an important role in describing single-particle properties. Taking $^{57}_{\Lambda_{c}}$Cu in Fig. \ref{Fig6-LcSingle_particle_energy.pdf} as an example, significant differences in the single-particle energy levels of the hyperon can be seen between the TW99-$\Lambda_{c}1$ and DD-LZ1-$\Lambda_{c}1$ models. For the hyperon potential, it is found that $V_{\sigma+\omega}$ and $V_{\rm{A}}$ given by these two models are basically consistent, while there is a significant difference in $V_{\rm{rea}}$. These studies demonstrate that a reasonable and effective treatment of nuclear medium effects is crucial for accurately describing both the bulk and single-particle properties of hypernuclei.

\begin{figure}
    \centering
    \includegraphics[width=0.48\textwidth]{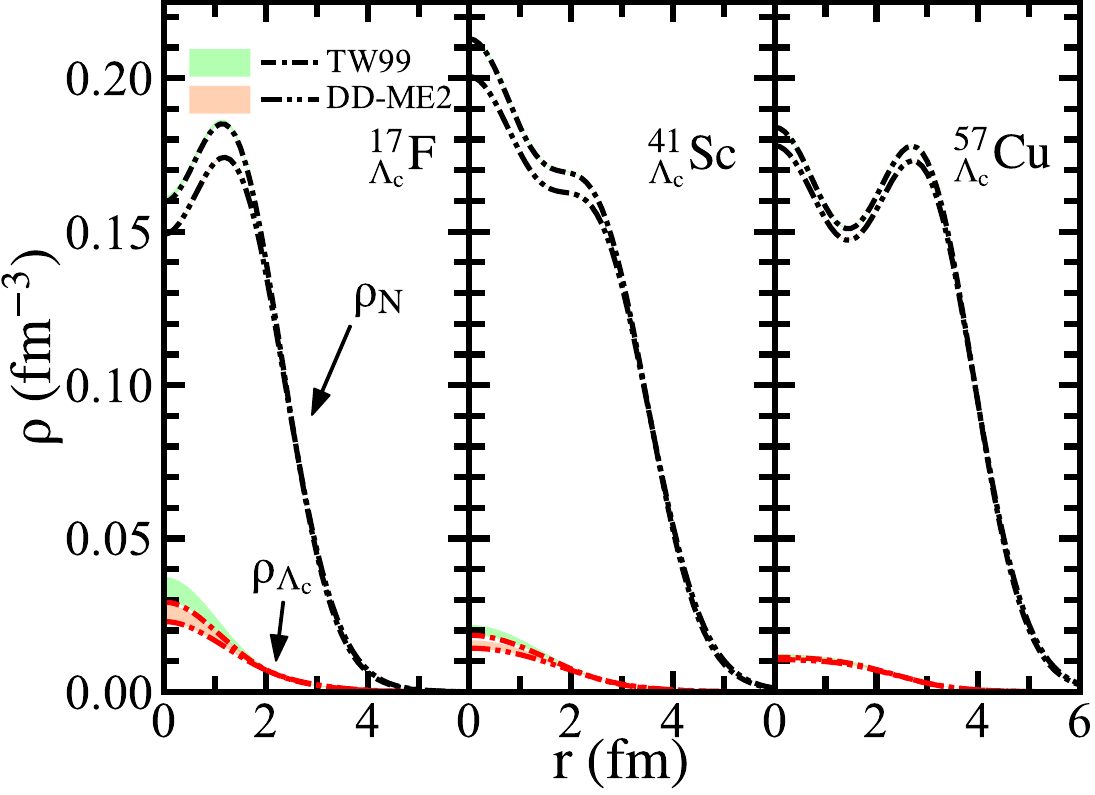}
    \caption{The nucleon (neutron and proton) and hyperon ($\Lambda_c^+$) densities in charmed hypernuclei $^{17}_{\Lambda_c}$F, $^{41}_{\Lambda_c}$Sc, $^{57}_{\Lambda_c}$Cu obtained by the DDRMF effective interactions TW99 and DD-ME2. The solid lines are derived from the $\Lambda_{c}$1 model, while the shaded areas represent results from $\Lambda_{c}$2.}
    \label{Fig8-Densities.pdf}
\end{figure}

In addition to affecting the single-particle energy levels of the $\Lambda_{c}^{+}$ hyperon, the model and uncertainties in $\Lambda_{c} N$ interactions also result in variations in the description of the bulk properties of hypernuclei. To illustrate this, the DDRMF models TW99 and DD-ME2, along with the effective $\Lambda_{c} N$ interaction, were used to determine the density distributions of hyperon and nucleons in $^{17}{\Lambda{c}}$F, $^{41}{\Lambda{c}}$Sc, and $^{57}{\Lambda{c}}$Cu, as shown in Fig. \ref{Fig8-Densities.pdf}. As the mass number of the hypernucleus increases, the nucleon density gradually extends outward. The strong interaction between nucleons and hyperon causes the hyperon density distribution to become more diffuse, resulting in a corresponding decrease in the central hyperon density. Furthermore, it can be observed that the central hyperon density predicted by DD-ME2 is consistently lower than that predicted by TW99. This implies that hyperon within the nucleus is more dispersed according to the predictions of DD-ME2, and a similar conclusion can be inferred from the nucleon density distributions. Additionally, for $ \Lambda_{c} $2, the uncertainty in hyperon densities also decreases when increasing the mass number. This reduction in uncertainty is mainly because $ \Lambda_{c}^{+} $ densities extend to larger radii, where the significance of the large potential difference at saturation density, as shown in Fig. \ref{Fig:Fig1-Pot_in_NM}, becomes less critical for heavier hypernuclei.

\begin{table*}[hbpt]
    \centering
    \caption{The binding energies $E$, single-particle energies $\varepsilon_{s.p.}$ of the $\Lambda_c^+1s_{1/2}$ state, matter radii $R_{m}$, hyperon radii $R_{\Lambda_c}$, and charge radii $R_{c}$ for the charmed hypernuclei with two DDRMF models TW99 and DD-ME2, accompanied by the results for their core of normal nuclei. The $\Lambda_{c}N$ effective interactions $\Lambda_c1$ are used. The radii are provided in unit of fm, and energies in MeV.}\label{Tab:Radius-bindingEnerige}
    \renewcommand{\arraystretch}{1.5}
    \setlength{\tabcolsep}{9pt}
    \begin{tabular}{cccccccccccc}
    	\hline\hline
    	&\multicolumn{5}{c}{TW99}&   & \multicolumn{5}{c}{DD-ME2}\\
    	         \cline{2-6}            \cline{8-12}
    	& $E$ & $\varepsilon_{s.p.}$ & $R_m$ & $R_{\Lambda_c}$ & $R_c$ & 
    	& $E$ & $\varepsilon_{s.p.}$ & $R_m$ & $R_{\Lambda_c}$ & $R_c$ \\\hline
    	$^{17}_{\Lambda_c}\mathrm{F}$  &-124.576&-2.151& 2.544 & 2.569 & 2.679 & &-129.302&-2.327& 2.599 & 2.686 & 2.728 \\
    	$^{16}\mathrm{O}$              &-123.286&      & 2.553 &       & 2.687 & &-127.750&      & 2.594 & & 2.727\\
    	\hline
    	$^{41}_{\Lambda_c}\mathrm{Sc}$ &-335.282&-1.543& 3.283 & 2.719 & 3.417 & &-344.087&-1.642& 3.336 & 2.873 & 3.467 \\
    	$^{40}\mathrm{Ca}$             &-333.993&      & 3.302 &       & 3.421 & &-342.608&      & 3.346 & & 3.464\\
    	\hline
    	$^{52}_{\Lambda_c}\mathrm{Cr}$ &-437.741&-2.104& 3.486 & 2.762 & 3.547 & &-444.951&-2.448& 3.529 & 2.837 & 3.589 \\
    	$^{51}\mathrm{V}$              &-435.795&      & 3.505 &       & 3.551 & &-442.589&      & 3.543 & & 3.588\\
    	\hline
    	$^{57}_{\Lambda_c}\mathrm{Cu}$ &-475.690&-0.943& 3.562 & 2.821 & 3.688 & &-481.975&-1.375& 3.601 & 2.873 & 3.727 \\
    	$^{56}\mathrm{Ni}$             &-474.894&      & 3.580 &       & 3.692 & &-480.682&      & 3.614 & & 3.727\\\hline\hline
    \end{tabular}
\end{table*}

As an extension, the binding energies of hypernuclei, single-particle energies of the $\Lambda_{c}^{+} 1s_{1/2}$ state, and their corresponding characteristic radii are provided based on TW99-$\Lambda{c}1$ and DD-ME2-$\Lambda{c}1$. The results are presented in Table \ref{Tab:Radius-bindingEnerige}. Considering the impurity effect induced by the $\Lambda_{c}^{+}$ hyperon, the corresponding results for nucleonic cores are also provided. Compared to their nucleonic cores, the introduction of the hyperon tends to reduce the matter radius of the hypernucleus. Comparing the results between the two models, it is evident that the characteristic radii predicted by DD-ME2-$\Lambda{c}1$ are generally larger than those predicted by TW99-$\Lambda{c}1$. This observation aligns with the conclusions drawn from the hyperon and nucleon density distributions shown in Fig. \ref{Fig8-Densities.pdf}. As the mass number increases, both sets of models indicate an increase in the matter radii of hypernuclei, while the radii of $\Lambda_{c}^{+}$ hyperon show significant differences. For TW99-$\Lambda_{c}1$, the hyperon radii increase gradually with the mass number, whereas for DD-ME2-$\Lambda_{c}1$, the hyperon radii increase initially and then remains almost constant. One possible explanation for this discrepancy is the differing balance between Coulomb repulsion and strong attractive interactions in the two models.

\section{Summary}\label{Summary and Outlook}

In summary, we extend the DDRMF theory to include the degrees of freedom of the $ \Lambda_{c}^{+} $ hyperon. The $ \Lambda_{c} N $ effective interaction is obtained by fitting the empirical hyperon potential of $ \Lambda_{c}^{+} $ in symmetric nuclear matter, based on first-principles calculations \cite{Haidenbauer.J-EPJA56(2020)195}. Due to the utilization of different cutoffs, the empirical hyperon potential of $ \Lambda_{c}^{+} $ is not uniquely determined and carries certain uncertainties. To mitigate the uncertainties arising from the fitting process in constructing the $ \Lambda_{c} N $ interaction, we have chosen the empirical hyperon potential with the minimum uncertainty as one of the fitting targets. Specifically, we selected the value at the Fermi momentum $ k_{F,n} = 1.05~\rm{fm}^{-1} $ and denoted the resulting interaction as $ \Lambda_{c}1 $. Considering our primary focus on describing the bulk and single-particle properties of $ \Lambda_{c}^{+} $ hypernuclei, where the nuclear medium density approaches saturation density, we also selected the empirical hyperon potential at saturation density, corresponding to a Fermi momentum of $ k_{F,n} = 1.35~\rm{fm}^{-1} $, as another fitting target, and the obtained interaction is named $ \Lambda_{c}2 $. Furthermore, we present the hyperon potential in symmetric nuclear matter, calculated based on the DDRMF Lagrangian and the $ \Lambda_{c} N $ effective interaction chosen in this work. We observed that the hyperon potentials derived from the DDRMF functionals DD-LZ1 and DD-ME$ \delta $ often exhibit significant differences, which reflect the uncertainties introduced by the DDRMF theory when describing the structures of $ \Lambda_{c}^{+} $ hypernuclei. In contrast, the hyperon potentials provided by TW99 and DD-ME2 show the smallest differences in the region below saturation density, potentially serving as rational models for accurately describing charmed hypernuclei. Regarding $ \Lambda_{c}1 $ and $ \Lambda_{c}2 $, the hyperon potentials obtained from the two sets of interactions display significant model dependence near saturation density and in the lower density region, respectively. Exploring these aspects can help us comprehend the impact of uncertainties in the hyperon potential at saturation and low densities on the description of hypernuclear structures.

Based on four typical DDRMF functionals, namely TW99, DD-ME2, DD-LZ1, and DD-ME$ \delta $, combined with the $ \Lambda_{c} N $ effective interaction, we explore the existence and stability of bound $ \Lambda_{c}^{+} $ hypernuclei. Since the hyperon is primarily distributed within the nucleus, where the nuclear medium density approaches saturation density, the bulk and single-particle properties of hypernuclei, are closely associated with interactions near saturation density. For $ \Lambda_{c}1 $, significant differences in the hyperon potential provided by various DDRMF functionals near the saturation density lead to markedly different conclusions regarding the existence of bound hypernuclei across different models. However, for the interaction $ \Lambda_{c}2 $, which is fitted to empirical hyperon potentials near saturation density, the uncertainty among the results of various DDRMF functionals is significantly reduced. To further clarify the effects influencing the existence and stability of $ \Lambda_{c}^{+} $ hypernuclei, we decompose the contributions of various components of the hyperon potential. Since the $ \Lambda_{c}^{+} $ particle is positively charged and isospin-zero, the hyperon potential can be decomposed into contributions from isoscalar mesons ($ \sigma $ and $ \omega^{\mu} $), photons, and rearrangement terms introduced by the density dependence of the meson-baryon coupling strengths. For the studied $ \Lambda_{c}^{+} $ hypernuclei, except for the light hypernucleus $ ^{5}_{\Lambda_{c}} $Li, the contribution from isoscalar mesons quickly saturates. Thus, the existence of hypernuclei depends on the contributions from Coulomb repulsion and the rearrangement term. In light hypernuclei, the impact of the rearrangement term is most significant. As the mass number increases, the contribution from Coulomb repulsion gradually becomes more dominant.

Furthermore, the bulk properties of $ \Lambda_{c}^{+} $ hypernuclei are presented using TW99-$ \Lambda_c1$ and DD-ME2-$ \Lambda_{c}1 $ effective interactions. It is found that the hyperon radius is more compact than the charge radius, and there is a reduction in the matter radius when compared to nucleonic cores. This suggests that the $ \Lambda_{c}^{+} $ hyperon is embedded within the hypernucleus. Moreover, the effect of introducing $ \Lambda_{c}^{+} $ hyperon on the nuclear charge radius exhibits variations between the two sets of effective interactions. Because the charmed hyperon carries an additional unit positive charge, it creates a competition between attraction from meson-nucleon interactions and repulsion from Coulomb interactions, affecting the charge distribution of nucleonic cores differently. Ultimately, the $ \Lambda_{c}^{+} $ radius tends to expand as the mass number increasing, which is primarily driven by the escalating influence of Coulomb repulsion. In addition to the $ \Lambda_{c} N $ interactions, the coupling channel $ \Lambda_{c} N $-$ \Sigma_{c} N $-$ \Sigma^{\ast}_{c} N $ has also been investigated in recent studies, which shows that the coupling channel is essential in $ \Lambda_{c} N $ bound states \cite{Liu.Yan.Rui-PRD85(2012)014015}. Consequently, comprehensive research on charmed hypernuclei systems is necessary, considering the mixing effects between nucleons and different charmed baryons. Correspondingly, given the current uncertainty in $ \Lambda_{c} N $ interaction research, it is hoped that more extensive theoretical and experimental research will be conducted in the future to refine the understanding of $ \Lambda_{c} N $ interaction.

\begin{acknowledgements}
This work was partly supported by the Fundamental Research Funds for the Central Universities, Lanzhou University (lzujbky-2022-sp02, lzujbky-2023-stlt01), the National Natural Science Foundation of China (11875152), the Strategic Priority Research Program of Chinese Academy of Sciences (XDB34000000).
\end{acknowledgements}

\end{document}